\documentclass{aa}  

\usepackage{graphicx}
\usepackage{txfonts}
\usepackage[colorlinks]{hyperref}
\hypersetup{
     colorlinks   = true,
     citecolor    = blue
}
\usepackage[utf8]{inputenc}
\usepackage{mathrsfs}
\usepackage{amssymb}
\usepackage{amsmath,bm}
\usepackage{physics}
\usepackage{relsize}
\usepackage{float}
\usepackage{textcomp}
\newcommand{\posc}{p_{\mathrm{osc}}}
\newcommand{\wdem}{W_{\mathrm{DEM}}}

\begin{document} 

   \title{Temporal evolution of oscillating coronal loops}

   \author{C. R. Goddard
          \inst{1}
          \and
          G. Nisticò
          \inst{2,3}}

   \institute{\inst{1}Max-Planck-Institut für Sonnensystemforschung, Justus-von-Liebig-Weg 3, 37077 Göttingen, Germany\\
              \email{goddard@mps.mpg.de}\\
              \inst{2}Institut für Astrophysik, Georg-August-Universität Göttingen, Friedrich-Hund-Platz 1, 37077 Göttingen, Germany\\
              \inst{3}Department of Physics, University of Calabria, Ponte P. Bucci, Cubo 31C, I-87036 Arcavacata di Rende, Italy}

   \date{Received January 2020 accepted April 2020}

  \abstract
   {Transverse oscillations of coronal structures are currently intensively studied to explore the associated magnetohydrodynamic wave physics and perform seismology of the local medium.}
   {We make a first attempt to measure the thermodynamic evolution of a sample of coronal loops that undergo decaying kink oscillations in response to an eruption in the corresponding active region.}
   {Using data from the six coronal wavelengths of SDO/AIA, we performed a differential emission measure (DEM) analysis of 15 coronal loops before, during, and after the eruption and oscillation.}
   {We find that the emission measure, temperature, and width  of the DEM distribution undergo significant variations on timescales relevant for the study of transverse oscillations. There are no clear collective trends of increases or decreases for the parameters we analysed. The strongest variations of the parameters occur during the initial perturbation of the loops, and the influence of background structures may also account for much of this variation.}
  {The DEM analysis of oscillating coronal loops in erupting active regions shows evidence of evolution on timescales important for the study of oscillations. Further work is needed to separate the various observational and physical mechanisms that may be responsible for the variations in temperature, DEM distribution width, and total emission measure.}

   \maketitle
\section{Introduction}

Transverse oscillations of coronal loops in erupting active regions have been intensively studied in recent decades. A recent and comprehensive review can be found in \cite{2019ASSL..458.....A}. These waves are considered to be global magnetohydrodynamic (MHD) kink eigenmodes, most commonly detected as the fundamental mode, with an antinode in the vicinity of the loop apex. The relation between the period of oscillation and the estimated loop length in \cite{2016A&A...585A.137G} and \cite{ 2019ApJS..241...31N} confirmed this interpretation.

Much of the interest in these oscillations stems from their seismological potential \citep[e.g.][]{2014SoPh..289.3233L} and the opportunity for detailed study of the associated MHD wave theory.
Recent examples of studies that attempted seismology include \cite{2015ApJ...799..151G, 2016A&A...589A.136P} and \cite{2019A&A...625A..35A}.
A further development is the detection of a  decayless  low-amplitude regime that is not associated with a flare or eruption \citep{2013A&A...552A..57N, 2015A&A...583A.136A}. \cite{2019ApJ...884L..40A} demonstrated that these waves may be used for seismology of quiet active regions. We also note that examples of apparently undamped, or even growing, high-amplitude oscillations associated with eruptive events have been reported \citep[e.g.][]{2012ApJ...751L..27W}.

As the observations and theory of these oscillations become increasingly complex, the uncertainties about the nature of the loops themselves become critical. \cite{2014LRSP...11....4R} recently reviewed loop observation and modelling.
There are different categories of loops, but for this study, only long, warm (1--2 MK), and non-flaring loops are important. Individual loop threads observed with the Atmospheric Imaging Assembly (AIA)  are normally part of larger bundles of similarly oriented threads. When such threads are analysed with higher resolution instruments \citep[e.g.][]{2013A&A...556A.104P, 2017ApJ...840....4A}, significantly increased fine structuring is not typically seen. However, the presence of further sub-resolution structuring (and the associated filling factor) of these individual apparently resolved threads is still a subject of debate, and is not considered here. \cite{2019ApJ...885....7K} recently confirmed that these long, warm loops show little evidence of expansion with height, nor do they show cross-sectional asymmetry, despite the expected expansion of the magnetic field with height. \cite{2019ApJ...873...26B} studied the emission around warm loops, finding that these loops are generally over-dense and largely isothermal, and are embedded in diffuse multi-thermal plasma with a peak temperature similar to that of the loops themselves. However, many studies have shown that the cross-field temperature structure of active regions loops varies from approximately isothermal to clearly multi-thermal \citep[e.g.][and references therein]{2011ApJ...731...49S, 2014ApJ...795..171S,2016ApJ...831..199S}.

Such individual loop threads are expected to be non-equilibrium structures, with lifetimes from tens of minutes to several hours predicted by thermodynamic analysis. In whole loop bundles, observation of thermal cycles are common \citep[see][and references therein]{2019arXiv191109710F}, with periods from 2--16 hours. Additionally, various plasma flows can be detected in a loop, which may result in a constant evolution. \cite{2018NatSR...8.4471S} reported density variations of the oscillating loop and related them to the period variation of the kink mode. Loop oscillation studies typically observe loops for about an hour, and a visible evolution of the intensity and structure is frequently observed \citep[see example time-distance (TD) maps in][]{2016A&A...585A.137G, 2019ApJS..241...31N}, and it is becoming increasingly important to analyse this evolution in greater detail.

Numerical simulations of transverse oscillations in radiatively cooling coronal loops were performed in \cite{2015A&A...582A.117M}, where a 20\% difference in amplitude after few oscillation cycles was found. Related analytical studies include those of \cite{2013SoPh..283..413A, 2017A&A...602A..50R} and \cite{2019A&A...625A..35A}. The role of the Kelvin Helmholtz instability (KHI)  during loop oscillations has been investigated by means of numerical simulation \citep[e.g.][]{2008ApJ...687L.115T, 2017ApJ...836..219A, 2019ApJ...876..100A} and can lead to the evolution of the density, temperature, and other observables \citep[e.g.][]{2016ApJ...830L..22A, 2018ApJ...863..167G}.

The aim of our study is to observationally explore the variation of the loop emission measure (and density by proxy), temperature, and width of the temperature distribution before, during, and after undergoing displacements and oscillations. In Section~\ref{sect:data_ana} the data and processing are described, in Section~\ref{sect:case} case studies are presented, collective results are reported in Section~\ref{sect:collect_ana}, and a discussion and conclusion are given in Section~\ref{sect:disc}. 

\section{Data analysis}
\label{sect:data_ana}

\begin{figure*}
\centering
\includegraphics[width=9.cm]{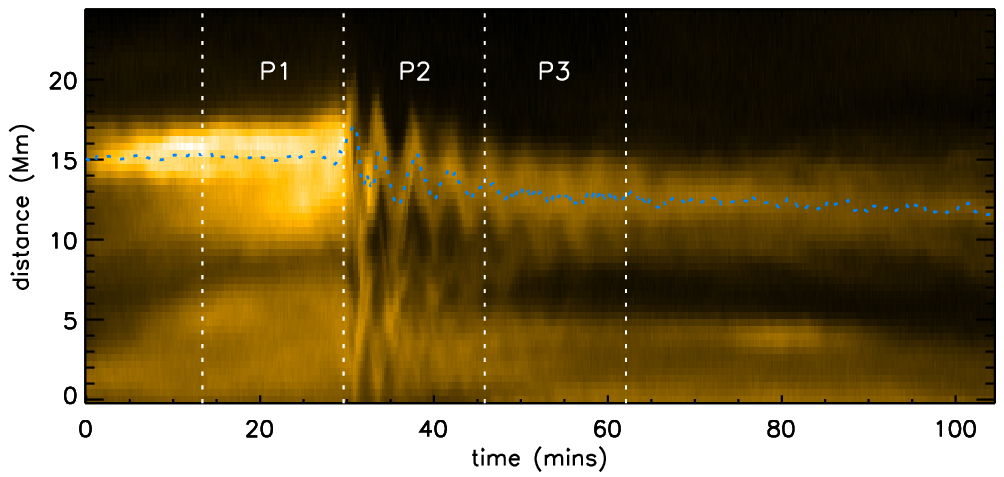}
\includegraphics[width=9.cm]{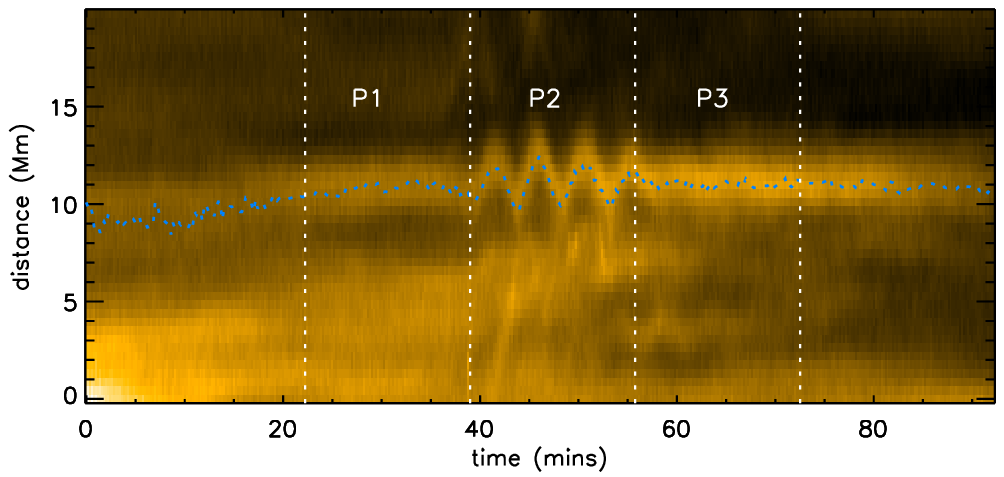}
\includegraphics[width=9.cm]{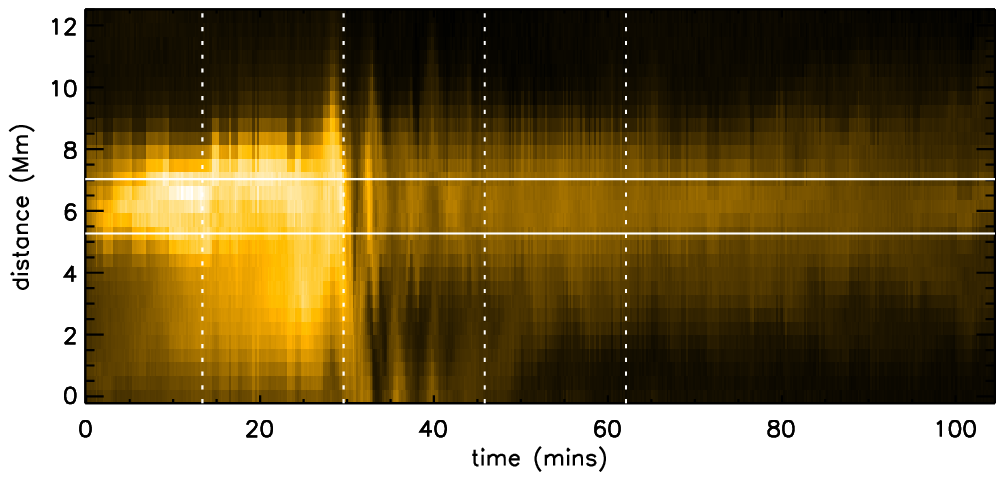}
\includegraphics[width=9.cm]{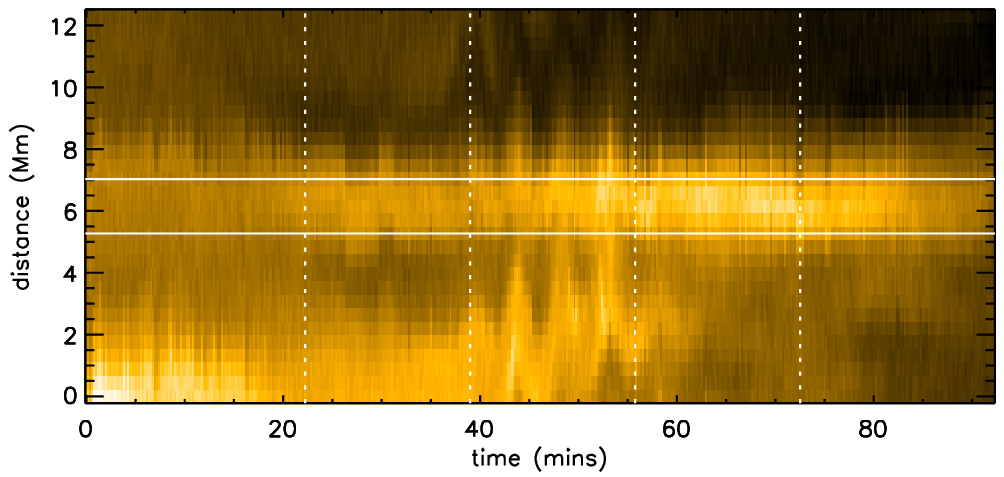}
\includegraphics[width=9.cm]{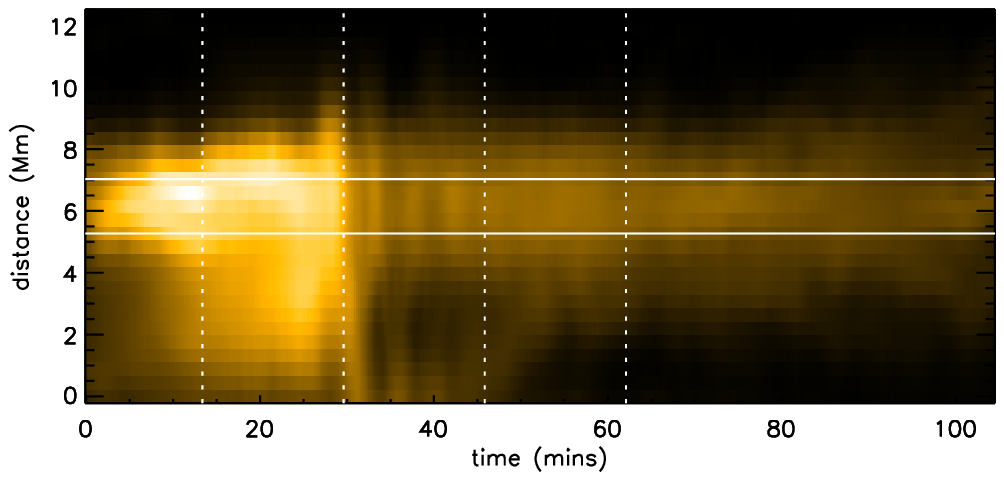}
\includegraphics[width=9.cm]{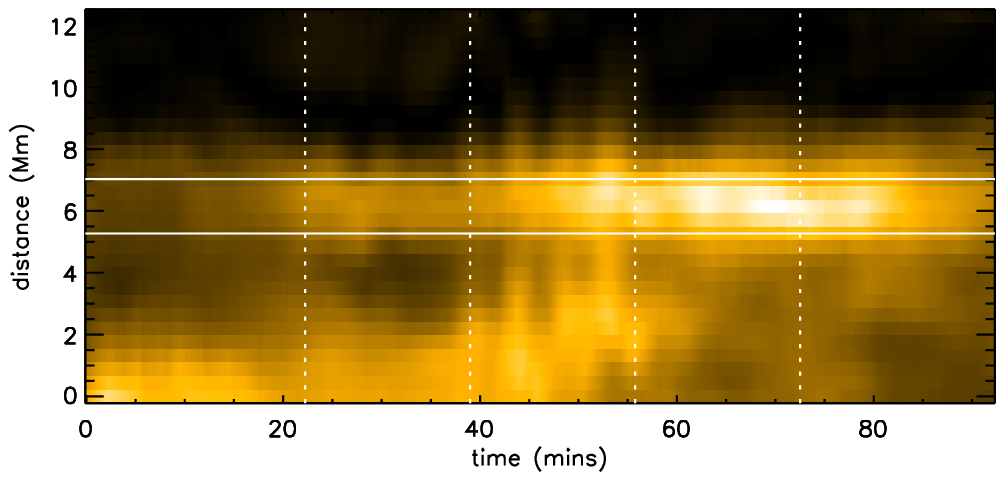}
\caption{\textit{Top:}  TD maps at 171 $\AA$ for event 28 loop 1 (left) and event 36 loop 2 (right). The dashed blue line is the oscillation time-series obtained from Gaussian tracking of the loop position. \textit{Middle:} Corresponding  detrended TD maps. The horizontal lines denote the averaging region, $\pm$ 2 pixels about the centre pixel. \textit{Bottom:} Same as above, but after smoothing in time with a three-minute boxcar, and subtracting the minimum intensity at each time.}
 \label{fig:td_maps}
\end{figure*}

\begin{figure*}
\centering
\includegraphics[width=6.cm]{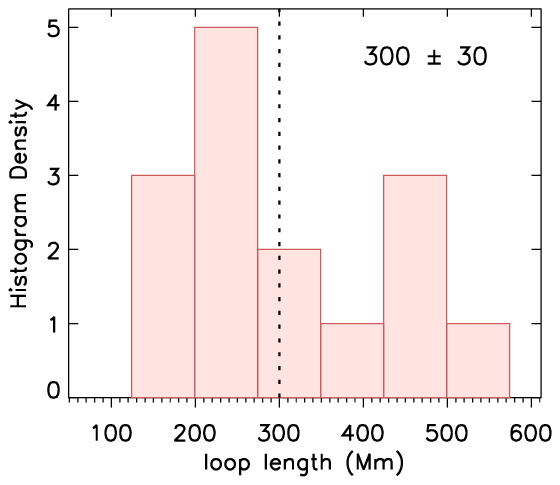}
\includegraphics[width=6.cm]{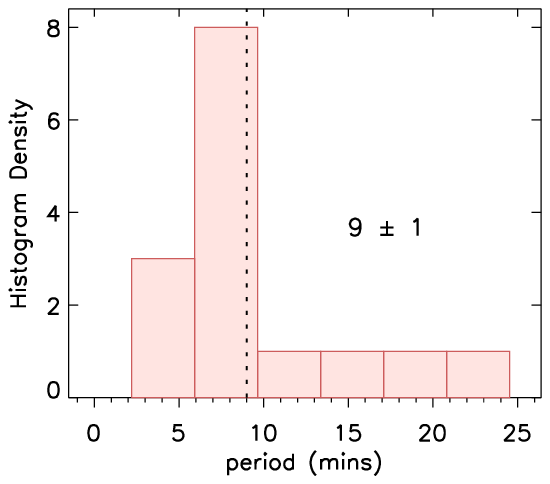}
\includegraphics[width=6.cm]{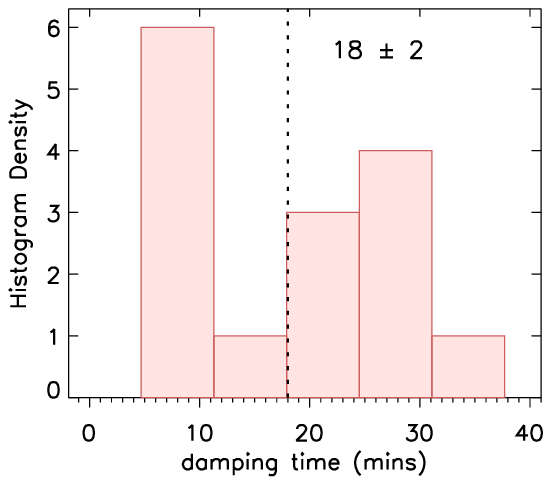}
\caption{Histograms of the loop length, oscillation period, and exponential damping time. The vertical dashed lines denote the mean values, labelled with the mean and its standard error.}
 \label{fig:osc_para}
\end{figure*}

\subsection{Event selection}
Kink oscillation events were selected from the catalogue given in \cite{2019ApJS..241...31N}, which were observed by the AIA on board the Solar Dynamics Observatory (SDO) \citep{2012SoPh..275...17L} from 2010--2018. The selection was based on the requirement for an isolated loop thread to be visible for the entirety of an oscillation (until the main oscillation has damped), and for a significant period before and after the oscillation (of the order of the damping time). A further requirement was that the loop position must be able to be tracked by Gaussian fitting of the transverse intensity profile to avoid manual tracking as in \cite{2016A&A...585A.137G} and \cite{2019ApJS..241...31N}, or more complex tracking procedures. 

Linear slits perpendicular to the loop axis were used for each event at 171 $\AA$ to create a series of TD maps, as described in \cite{2019ApJS..241...31N}, and the clearest TD map was confirmed to meet the above criteria. This procedure resulted in just 15 events for further analysis. The next two subsections outline the procedure we performed for each oscillation event.

We note that our definition of the oscillating loop is an individually distinguishable, apparently monolithic \lq thread~seen at 171~$\AA$, that is, a warm loop ($\sim$ 1 MK) with a lifetime of about an hour or longer. This is in contrast to long-lived  loop systems, which may be analysed for tens of hours \citep[e.g.][]{2016ApJ...827..152A, 2019arXiv191109710F}.

Our analysis is limited to a few pixels around the apparent loop centre determined in the 171 \AA~images to attempt a first search for any evolutionary signatures in an ensemble of oscillating loops. We do not attempt to characterise the transverse temperature or density profiles of the structures \citep[e.g.][]{2017A&A...600L...7P,2018ApJ...863..167G}. This would require a careful treatment of the background in all wavelengths, which may be a natural extension of the work in the future.

\subsection{Initial processing}
\label{sect:init_process}
Data for all coronal filters of AIA (94, 131, 171, 193, 211, and 335 \AA) were downloaded for a time window before and after the reported oscillation start time. This was done using the SolarSoftWare (SSW) function \texttt{ssw\_cutout\_service}, and included frames with the \texttt{AEC = 1} flag to retain the 12-second cadence for each wavelength. The data cubes for each wavelength were then processed with \texttt{read\_sdo} and \texttt{aia\_prep}. For the chosen slit position, TD maps were created at each wavelength (co-aligned based on the header information), and averaged over a 5-pixel width perpendicular to the slit to reduce noise. The selected events are listed in Table~\ref{tab:1}. All analysed loops were off-limb, except for event 29, loop 2.

An oscillation time-series was then obtained by fitting the transverse intensity profile of the loop at 171 \AA~with a Gaussian plus a second-order polynomial at each time, taking the peak position as the loop centre, as in \cite{2016A&A...585L...6P}. This is taken to be the loop centre position at all other wavelengths, regardless of whether the loop is visible. The top panels of Fig.~\ref{fig:td_maps} show examples of a loop-centre track overplotted on the 171 \AA~TD maps (the rest are presented in Appendix~\ref{app:1}).  Detrended ~TD maps at each wavelength were then created by extracting $\pm$14 pixels about the loop centre position (29 pixels in total). 

These detrended TD maps were smoothed in time with a boxcar of 15 time steps (3 minutes) to reduce noise because the DEM analysis is sensitive to small changes in the channel intensities (and their errors). The errors on the smoothed TD map intensities were estimated using the SSW routine \texttt{aia\_bp\_estimate\_error}. A limited form of background subtraction was then performed for each detrended and smoothed TD map by subtracting the minimum intensity at each time. This was performed separately for each wavelength and removed some of the influence of the background intensity and its variation over time. The middle and bottom panels of Fig.~\ref{fig:td_maps} show two examples of detrended and smoothed TD maps at 171 \AA. Averaging to obtain time-series of all parameters was performed within $\pm$ 2 pixels about the central pixel.

Because different slit positions were sometimes used with respect to the catalogue, the oscillation period ($\posc$) and exponential damping time ($\tau_d$) were estimated again for each event \citep[following][]{2016A&A...585A.137G}, and are given in Table~\ref{tab:1}. Figure~\ref{fig:osc_para} shows histograms and the mean values of the loop length, period, and damping time. A more detailed treatment of the damping profiles is not required \citep[e.g.][]{2016A&A...585L...6P, 2016A&A...593A..59M} because the damping time is just used to define consistent time chunks for the further collective analysis of all events. The time of the initial perturbation was determined ($t_0$), and then three temporal windows were defined, $\mathrm{P1}=[t_0 - 2\tau_d, t_0]$, $\mathrm{P2}=[t_0, t_0 + 2\tau_d]$, and $\mathrm{P3}=[t_0 + 2\tau_d, t_0 + 4\tau_d]$. These three time intervals were found to define the  pre-oscillation~ (P1),  oscillation~ (P2), and  post-oscillation~ (P3) phases for most events. In some cases, marked by an asterisk in Table~\ref{tab:1}, intervals of just one damping time were used. When these intervals extended slightly beyond the limits of the TD map, this resulted in correspondingly shorter P1 or P3 averaging intervals.

\begin{table}[]
\caption{Fifteen loop oscillation events chosen for analysis. The first two columns refer to the event and loop IDs given in \cite{2019ApJS..241...31N}, and the loop length is taken from the catalogue. The period and damping time are determined by the same method as in \cite{2016A&A...585A.137G}, using the present data. Events with asterisks note cases where temporal windows of just one damping time were used in the further analysis.}
\begin{tabular}{c|c|c|c|c}
\hline
Event ID & Loop ID & Length & Period & Damping Time \\ 
& & (Mm) & (min) & (min) \\ \hline
3        & 1       & 213         & 2.2    &  4.7     \\
5        & 1       & 438         & 8.6    & 11.2    \\
19       & 1       & 499         & 16.8   & 26.1    \\
22       & 1       & 432         & 18.4   & 31.1    \\
27       & 1       & 162         &  7.6   & 19.6    \\
28       & 1       & 234         &  4.0   &  8.1     \\
29       & 2       & 407         &  8.5   & 22*     \\
36       & 2       & 347         &  6.7   & 10.1    \\
36       & 10      & 238         &  6.0   & 10.7    \\
37       & 4       & 222         &  4.8   &  8.4     \\
63       & 2       & 183         &  6.0   & 24.4*   \\
64       & 6       & 431         & 20.8   & 25.9    \\
72       & 3       & 124         &  6.6   & 30.3*   \\
90       & 1       & 254         & 10.5   & 29*     \\
93       & 1       & 326         &  8.2   & 13.7   \\
\hline
\end{tabular}
\label{tab:1}
\end{table}

\subsection{Differential emission measure analysis}

\begin{figure*}[htpb]
    \centering
    \includegraphics{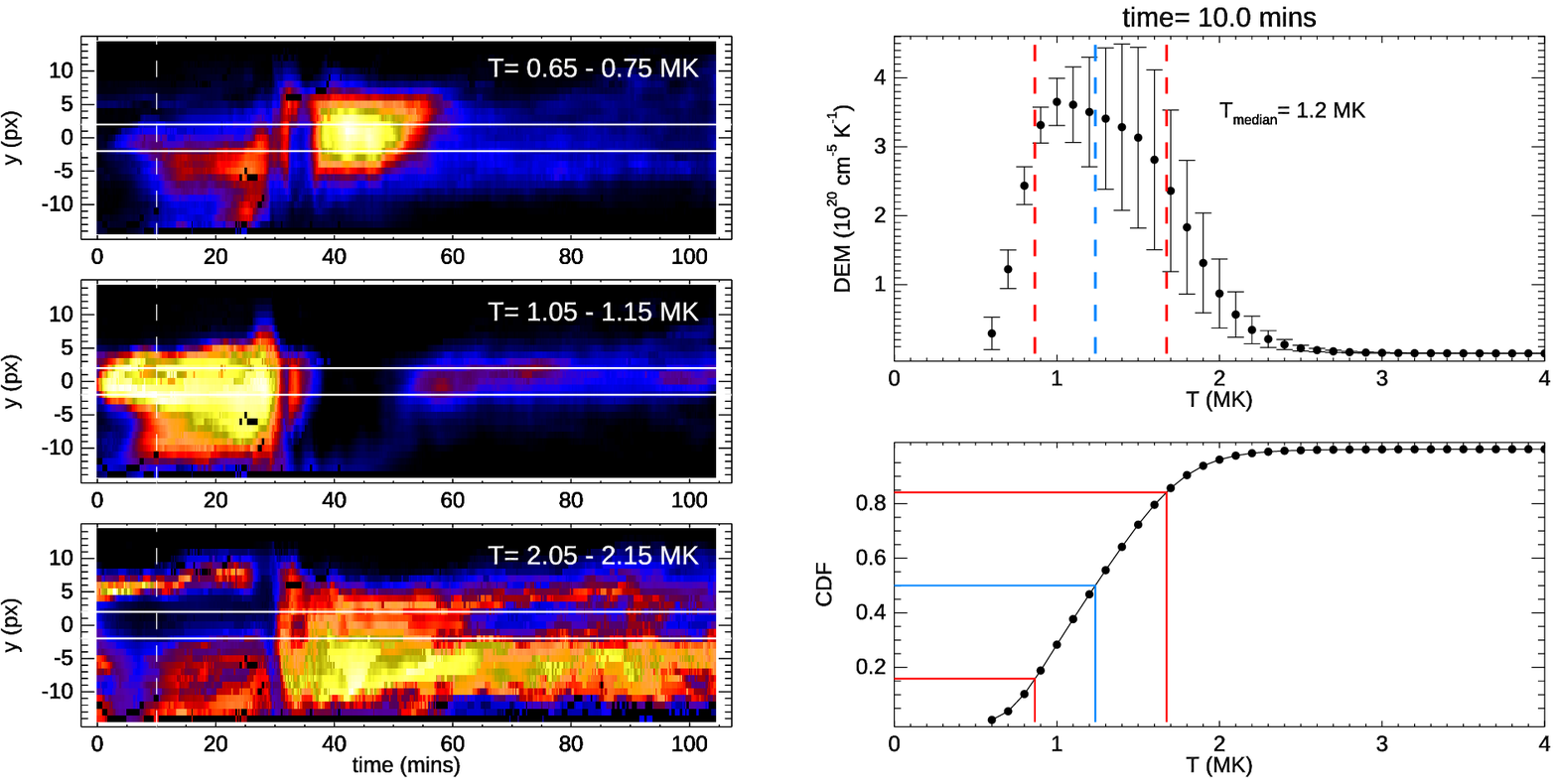}
    \caption{Left three panels: DEM maps for three of the temperature bins, as labelled. The two horizontal white lines denote the central five pixels, which are averaged over at each time. The right panels show the distribution of $DEM$ and $CDF$ with temperature at 10 min (overplotted vertical white line in the left panels). The blue lines represent $T_\mathrm{median}$ , and the red lines represent $T_\mathrm{min,max}$.}
    \label{fig:DEM_EM_temp}
\end{figure*}

The light recorded by extreme ultraviolet (EUV) telescopes (e.g. STEREO/EUVI and SDO/AIA) is sensitive to the square of the electron plasma density ($n_\mathrm{e}$) distributed along the line of sight (LOS, $z$). The amount of plasma along the LOS is defined as the emission measure,

\begin{equation}
EM = \int n_e^2 dz. 
\end{equation}

Because the plasma integrated along the LOS can be at different temperatures, it is useful to define the differential emission measure $dEM/dT$ (more commonly indicated as $DEM$). Therefore, the associated emission measure, $EM$, is given as an integral of $DEM$ over the temperature, $T$, as 
\begin{equation}
      EM = \int n_\mathrm{e}^2 \frac{dz}{dT} dT = \int \frac{dEM(T)}{dT} dT = \int DEM(T) dT. 
\end{equation}

The temperature-response function ($K$) of the instrument for a given waveband ($\lambda$) filters the observed plasma in a certain temperature interval. Therefore, the associated flux or intensity is given by 
\begin{equation}
    I_\lambda = \int {K_\lambda(T) DEM(T) dT}.
\end{equation}
The response functions of EUV imagers have relatively narrow-band peaks, but include spectral contributions from a wide range of temperatures \citep[e.g.][]{2012SoPh..275...17L}. Moreover, only a limited set of observations in different EUV wavebands are available (e.g. the six channels of AIA), and the inversion of the corresponding intensities into a DEM distribution constitutes an ill-posed problem  \citep[e.g.][]{Hannah2012}.
 Several algorithms have been developed for the determination of the DEM distribution of coronal structures observed in EUV and X-ray wavebands. For example, \citet{Aschwanden2008} forward-modelled the DEM distribution with Gaussian functions. Other methods are based on regularisation to avoid calculating negative DEMs \citep{Hannah2012, Plowman2013}, or they are based on the concept of sparsity \citep{Cheung2015}. Recently, multiple methods were compared in \cite{2019SoPh..294..135M}. For this study we adopted the DEM analysis for the AIA images developed by \citet{Hannah2012}. The code is available on-line  \footnote{\url{https://github.com/ianan/demreg}}. 
 
The DEM code was run over the detrended and background-subtracted intensity TD maps at the six coronal EUV channels of AIA (we excluded the 304 $\AA$ wavelength). These maps were given as input, along with the error estimates (obtained from \texttt{aia\_bp\_error\_estimate\_error}) and the response functions at the time of observation (determined using \texttt{aia\_get\_response}). The DEM code returns a series of DEM maps calculated over user-defined temperature intervals. We chose a temperature range spanning 0.6 to 10 MK, defined over 95 equispaced bins of 0.1 MK.

The left panels in Figure \ref{fig:DEM_EM_temp} show the DEM maps for event 28, loop 1 at the temperature intervals centred on 0.7, 1.1, and 2.1 MK. We averaged the DEMs over the indicated central pixels for each time and built a DEM distribution as a function of the temperature. An example is given in the top right panel of Fig.~\ref{fig:DEM_EM_temp}, where we show the DEM distribution for a specific time (t=10 min). Most of the DEM contribution is located in the temperature range 0.8 -- 1.5 MK, which is the typical temperature range for non-flaring warm coronal loops. 

The typical shape of the DEM distributions is not purely Gaussian. The obtained distributions  are ofen asymmetric, and they are sometimes double-peaked, probably due to the presence of multiple structures along the LOS. Therefore the temperature of the DEM peak is not a good observable for the  characteristic~temperature of the DEM distribution and does not represent its average. We consider the median temperature of the DEM distribution as a good observable. This was calculated by considering the cumulative distribution function ($CDF$) of the DEM distribution for any time as
\begin{equation}
    CDF(T_i) = \frac{\sum^i_{j=1} DEM_j \Delta T_j}{EM}
    \label{eq:cdf}
,\end{equation}
with $EM = \Sigma_k DEM_k \Delta T_k$. Equation \ref{eq:cdf} has the same meaning as a probability, and the sum over all the temperature bins returns 1, that is, the normalised total area of the DEM distribution. The median of the temperature divides the distribution into two sectors of equal area. Therefore we calculated the median temperature by linear interpolation as the value $T_\mathrm{med}$ where $CDF(T_{\mathrm{med}} )=0.5$ (we refer to it as $T$ below). The standard deviation is defined as the temperature interval around the median, with a probability of 0.68. To calculate the extrema of this interval, we therefore determined the lower and maximum temperature as $CDF(T_{\mathrm{min}} )=0.16$ and $CDF(T_{\mathrm{min}} )=0.84$. Because of the asymmetry of the distribution, the values $T_\mathrm{min,max}$ are in general not equidistant from $T_{\mathrm{med}}$. The time variation of the standard deviation is related to the changes in the broadness of the DEM distribution, which has important physical implications. We therefore considered another observable in our analysis: the width of the DEM distribution as $\wdem = \abs{T_{\mathrm{max}}-T_\mathrm{min}}$, along with the total emission measure, $EM$, and median temperature, $T$. These definitions are shown in the right panels of Fig.~\ref{fig:DEM_EM_temp}.

\begin{figure*}
\centering
\includegraphics[width=9.cm]{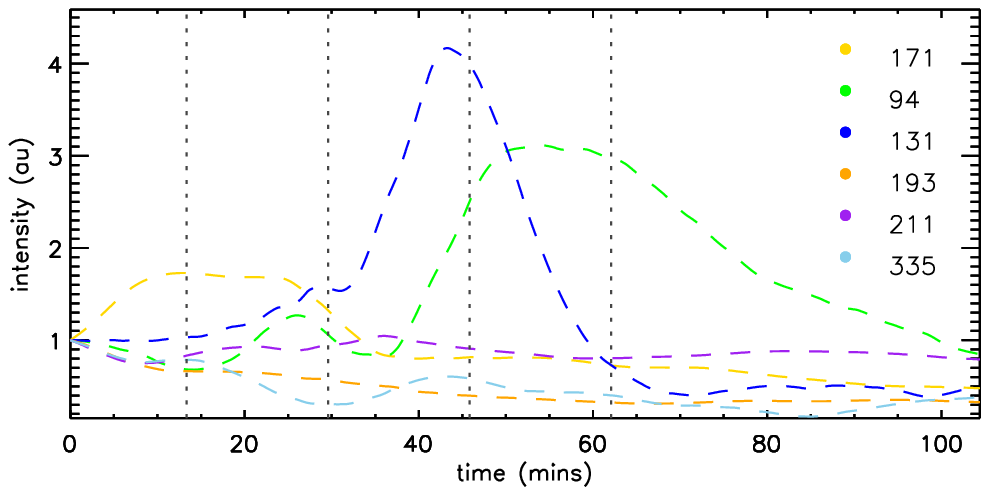}
\includegraphics[width=9.cm]{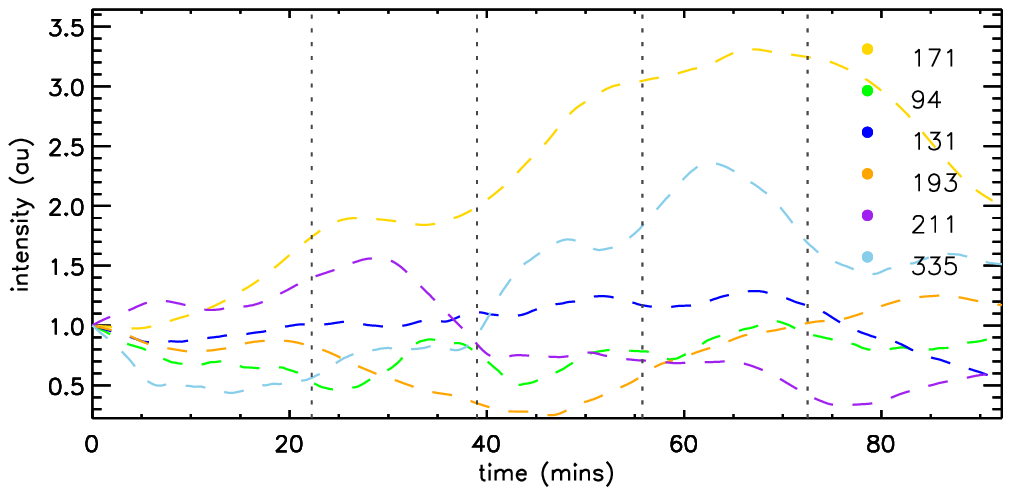}
\includegraphics[width=9.cm]{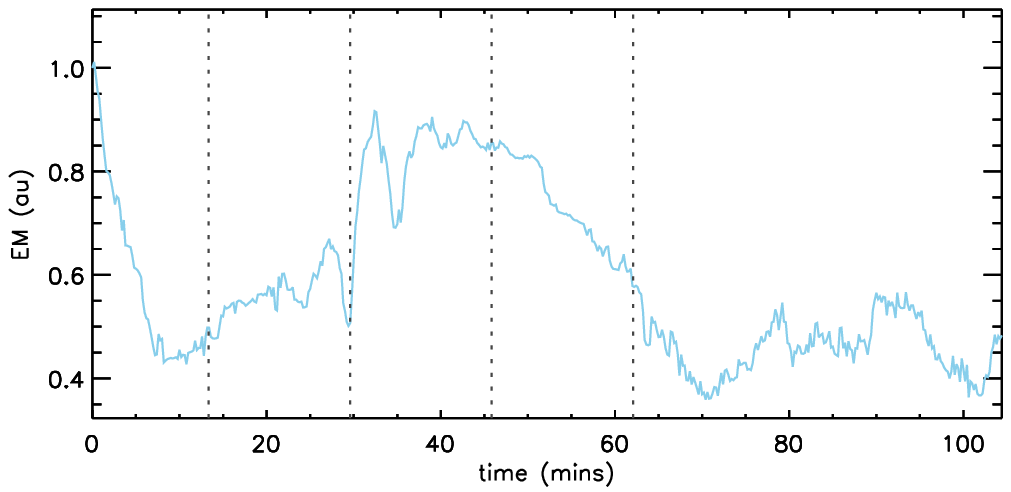}
\includegraphics[width=9.cm]{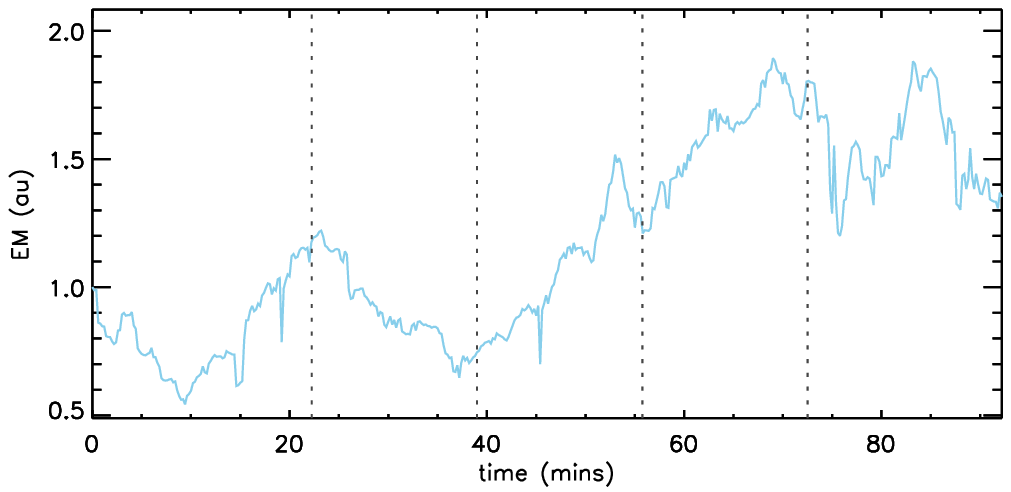}
\includegraphics[width=9.cm]{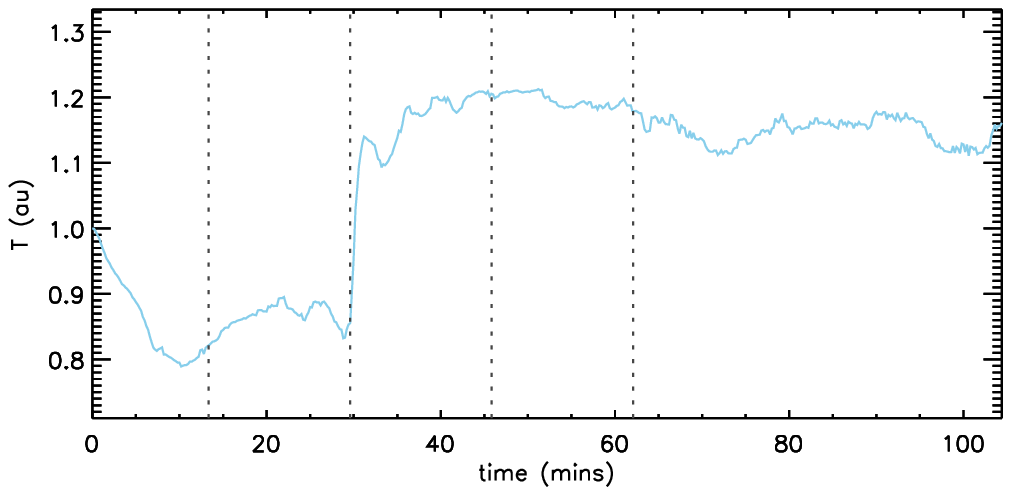}
\includegraphics[width=9.cm]{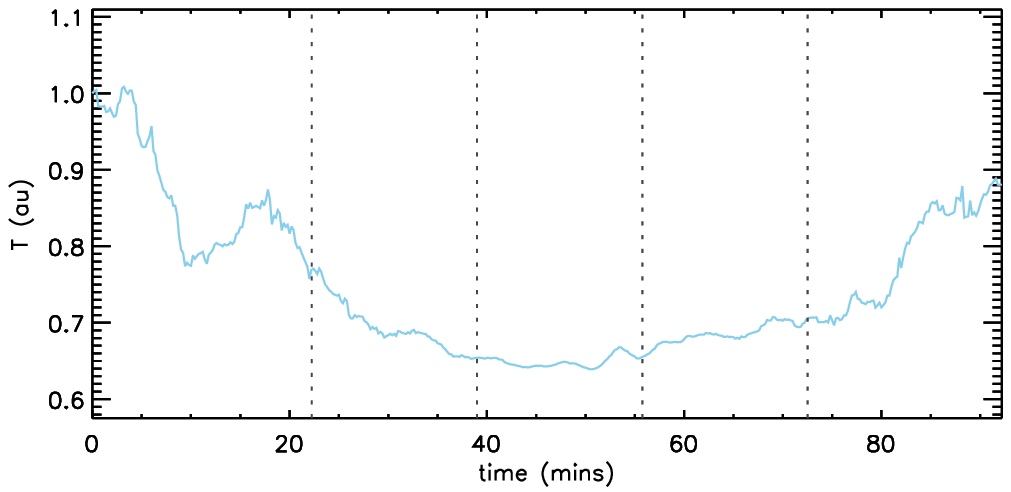}
\includegraphics[width=9.cm]{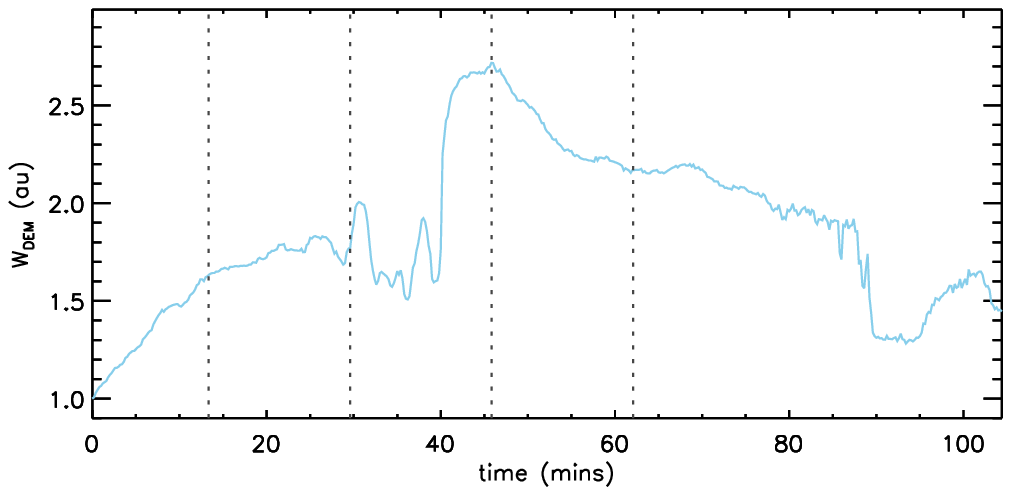}
\includegraphics[width=9.cm]{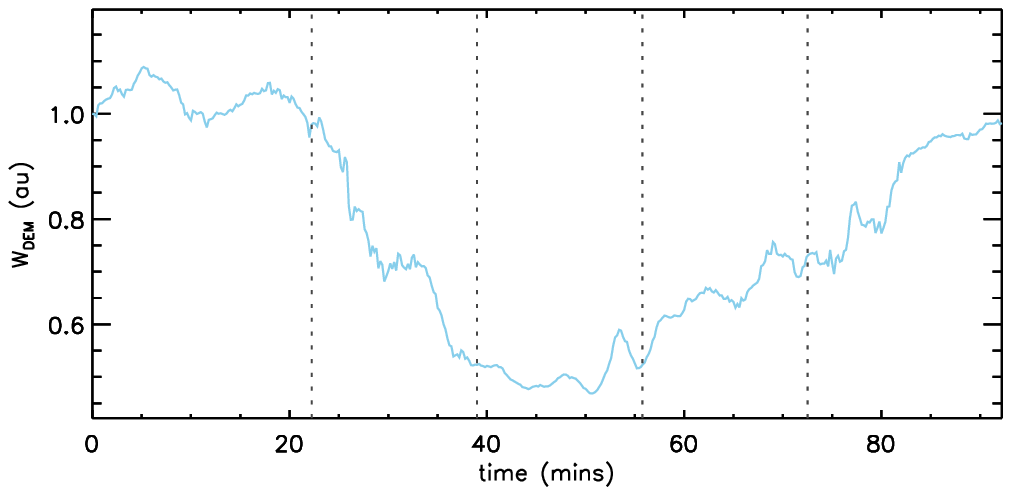}
\caption{ \textit{Top row}: Smoothed normalised intensity time-series for the six coronal wavelengths. \textit{Bottom three rows}: Normalised time series of $EM$, $T,$ and $\wdem$. In all panels the vertical dashed lines enclose the three time intervals P1, P2, and P3 (as in Fig.~\ref{fig:td_maps}). The left plots are for event 28, loop 1, and the right plots for event 37, loop 4.}
 \label{fig:case1}
\end{figure*}

From the DEM maps we created time series of $EM$, $T$ and $\wdem$ averaged within the central 5 pixels for each time. Examples of the final time series are shown in the bottom three rows of Fig.~\ref{fig:case1}, which are normalised by the initial values. 

\section{Case studies}
\label{sect:case}
Two case studies are now presented to show the analysis of the TD maps and the observables obtained from the DEM analysis. This also allows for qualitative discussion of the trends for the emission measure, the temperature, and the broadness of the DEM distribution. 

\subsection{Event 28, loop 1}

This loop oscillation was previously analysed in a seismological manner as  loop 3~in \cite{2016A&A...589A.136P}, where a context image of the loop and active region is shown, and in \citet{Nistico2013a} with the first detection of low-amplitude decayless kink oscillations. The top left panel of Fig.~\ref{fig:td_maps} shows the TD map at 171 $\AA$, with the fitted loop centre position overplotted. The first two dashed white lines enclose the time chunk P1, the second and third lines enclose chunk P2, and the third and fourth lines enclose chunk P3, as labelled. The middle left panel shows the  detrended~TD map, and the smoothed version is shown in the bottom panel.

Plots related to the DEM decomposition of this event are shown as an example in Fig.~\ref{fig:DEM_EM_temp}. The DEM map for 1.05-1.15 MK clearly shows the loop plasma seen in the detrended 171 \AA~TD map. The cooler and hotter DEM maps show the appearance of other structures that are co-spatial with the loop plasma at about minute 30, highlighting the difficulty and limitations of the analysis. Additionally, the loop is seen as a reduction in the DEM in the higher temperature bin. In the top left panel of Fig.~\ref{fig:case1} the intensity time series for each wavelength are plotted (after secondary smoothing with a boxcar of 10 minutes for clarity), normalised by the starting values. The remaining three left panels show the $EM$, $T,$ and $\wdem$ normalised to the starting values. The starting values were $EM = 8.5\times 10^{26} \mathrm{cm}^{-5}$, $T = 1.6 \,\mathrm{MK,}$ and $\wdem = 0.5 \,\mathrm{MK}$.

The second dashed line in all plots represents the onset of the initial perturbation of the loop ($t_0$), the start of P2. Most of the damping then occurs during P2. Some residual oscillation plus the decayless oscillations are visible during P3. For this event, the loop remains clearly visible for a significant duration after P3. This is not typical in the 15 events, however, and so the collective analysis is restricted to the three time intervals defined based on the estimated damping time.

The intensity at 171 $\AA$ increases over the first 10 minutes and reaches a plateau before it decreases after the loop is perturbed by an eruption. The $T$ time series  shows that this corresponds to an $\approx 20 \%$ increase in the measured temperature. $\wdem$ also increases slightly later, but began to increase from the start of the observation. The $EM$ time series is dominated by a more gradual rise and fall, which is due to the plasma, which appears at cooler and hotter temperatures. \citet{Nistico2013a} described an expansion or rise of the loops just before the eruption, which may be a signature of gradual evolution occurring in the loop. The intensity peaks at 94 and 131 $\AA$ that occur after the main part of the oscillation are due to the passing through the slit of multi-thermal plasma associated with the eruption. Such a transient feature cannot be easily removed from the analysis.

During the oscillation, a modulation of $EM$ with approximately $\posc$ is present, which is likely due to a variation in column depth, but might be a combination of multiple factors \citep[e.g.][]{2016ApJS..223...24Y}. Modulation of the temperature can also be seen, which might be due to higher temperature structure in the background that the loop may periodically pass over. 

After P3, $EM$ and $T$ remain almost constant, therefore the loop and surrounding environment appear to have reached a new equilibrium following the eruption and oscillation. However, all parameters varied strongly before P1. It is almost impossible to separate the components that are due to the eruption and flare, and those due to the oscillation. However, we can probably rule out an overall heating or cooling cycle of the loop because the DEM parameters and intensities are almost constant after P3. 

\subsection{Event 37, loop 4}

This loop oscillation was previously analysed in a seismological manner as  loop 1~in \cite{2016A&A...589A.136P}, where a context image of the loop and active region is shown. The TD maps are shown in the right panel of Fig.~\ref{fig:td_maps}, and the intensity, $EM$, $T,$ and $\wdem$ time series in the right panels of Fig. \ref{fig:case1}, as described above. The starting values of the parameters used for normalisation were $EM = 0.71\times 10^{26} \mathrm{cm}^{-5}$, $T = 1.3\, \mathrm{MK,}$ and $\wdem = 0.80\, \mathrm{MK}$.

Here we also see some residual oscillation during P3. However, for these two events, the chosen time intervals adequately represent the phases; prior to the large amplitude oscillations (and eruption), during the oscillation (post eruption), and after the main oscillation phase.

In contrast to the previous event, no jumps in the intensity at 94 and 131 $\AA$ during P2 (following the eruption) are observed. The intensity at 171 $\AA$ increases gradually throughout the observations before it drops again after P3. The brightening of the oscillating loop is clearly seen in the original TD maps. The $EM$ time series follows this trend, meaning that the quantity of low-temperature plasma increases. It is unclear, however, whether this is due to the increasing loop density or to movement of the second loop nearby in the 171 $\AA$ TD maps. The $T$ and $\wdem$ time series show a gradual evolution that begins well before the eruption, and so the increase in the $EM$ is likely to be due to evolution of the loop resulting from general evolution of the active region during the eruptive event, rather than directly due to the oscillation.

\section{Collective analysis}
\label{sect:collect_ana}

\begin{figure*}
\centering
\includegraphics[width=6.cm]{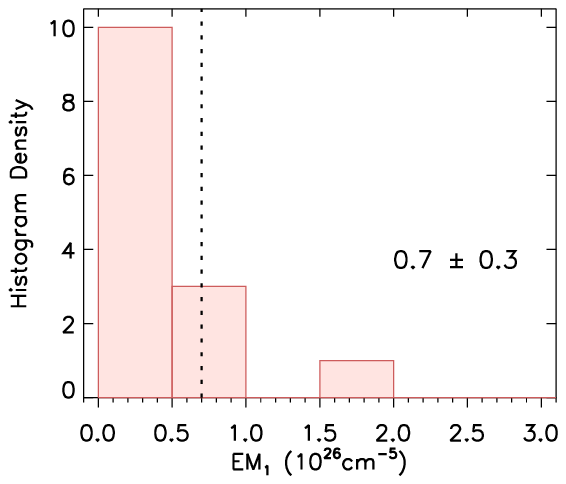}
\includegraphics[width=6.cm]{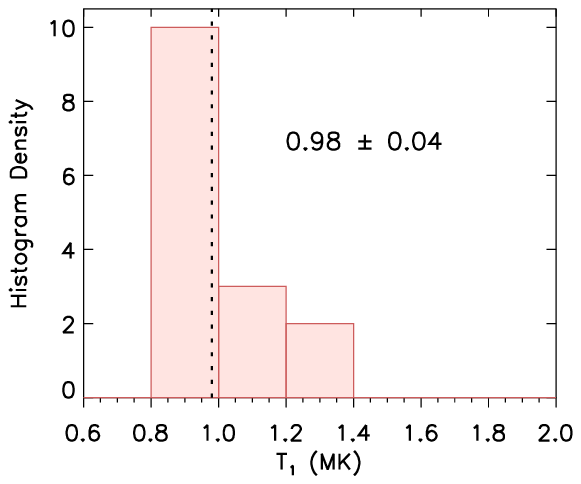}
\includegraphics[width=6.cm]{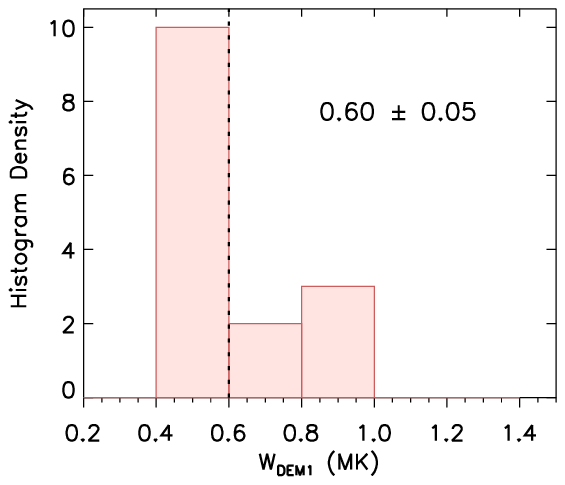}
\includegraphics[width=6.cm]{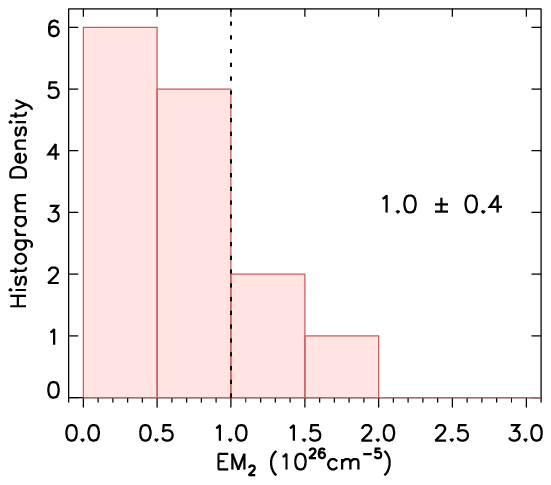}
\includegraphics[width=6.cm]{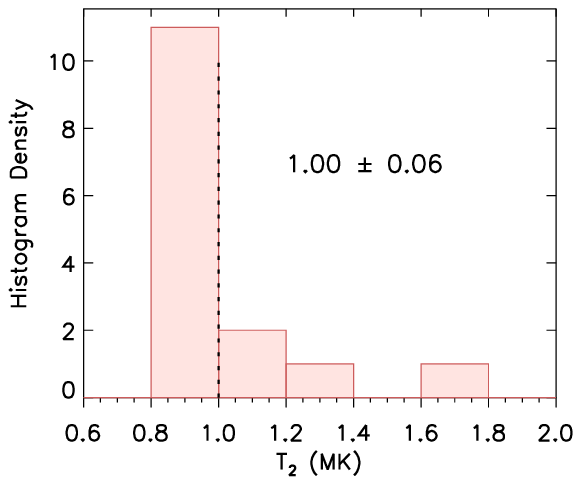}
\includegraphics[width=6.cm]{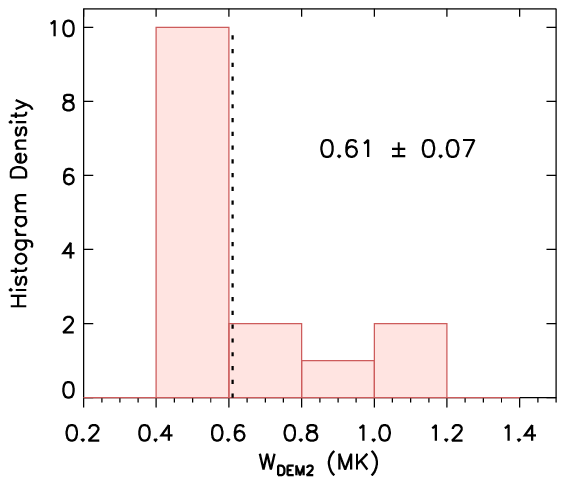}
\includegraphics[width=6.cm]{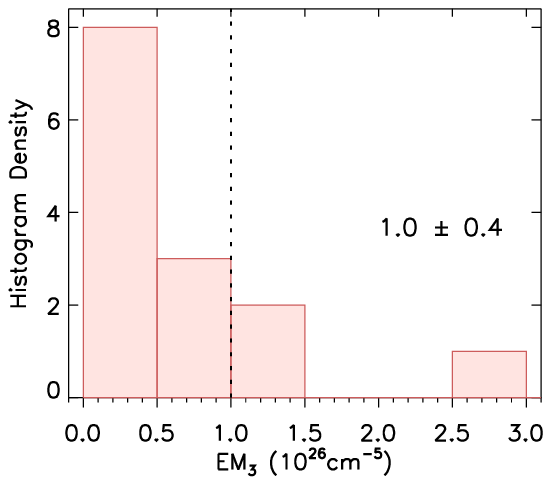}
\includegraphics[width=6.cm]{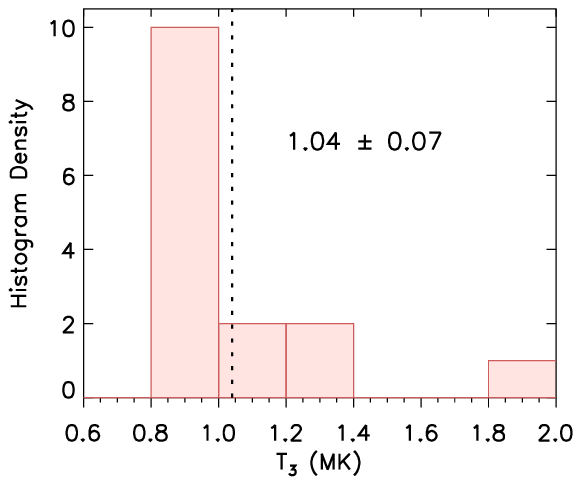}
\includegraphics[width=6.cm]{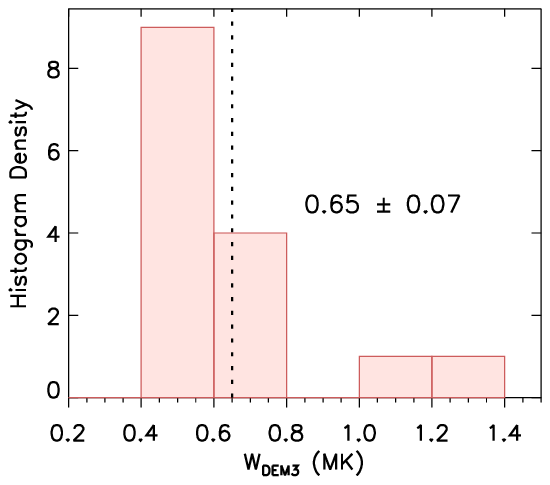}
\caption{Histograms of $EM$, $T,$ and $\wdem$, left to right, in each temporal window (P1, P2, and P3, top to bottom). The vertical dashed lines denote the mean values, as given in Table.~\ref{tab:2}. The values for some events lie outside of the plotted ranges.}
 \label{fig:windows_para}
\end{figure*}

\begin{figure*}
\centering
\includegraphics[width=6.cm]{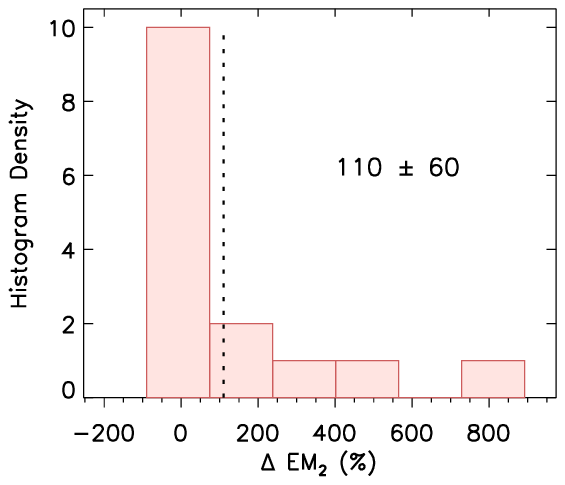}
\includegraphics[width=6.cm]{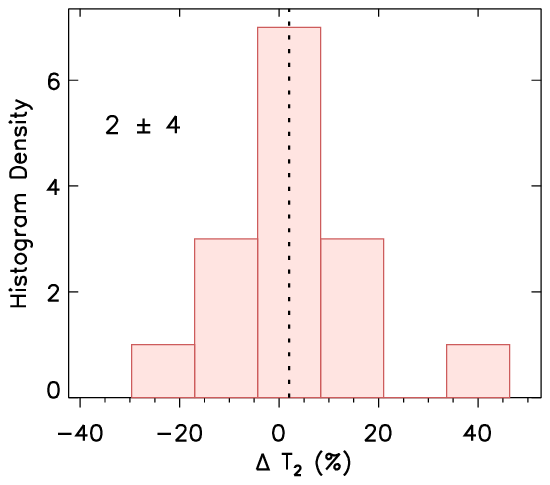}
\includegraphics[width=6.cm]{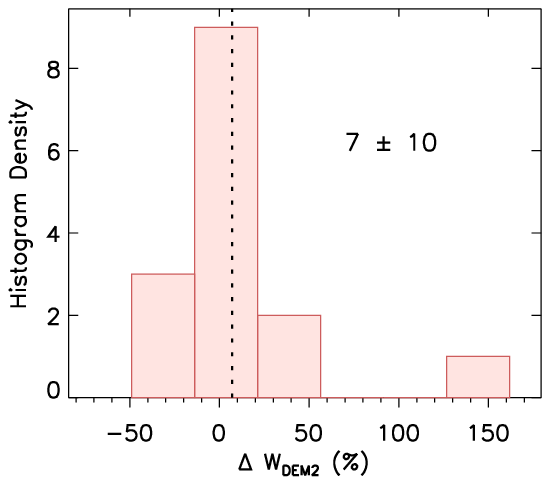}
\includegraphics[width=6.cm]{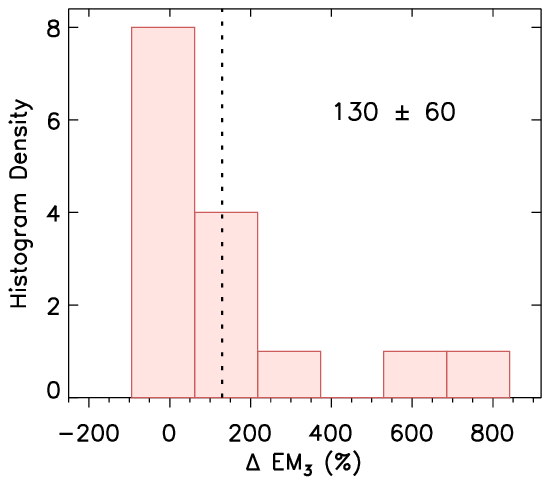}
\includegraphics[width=6.cm]{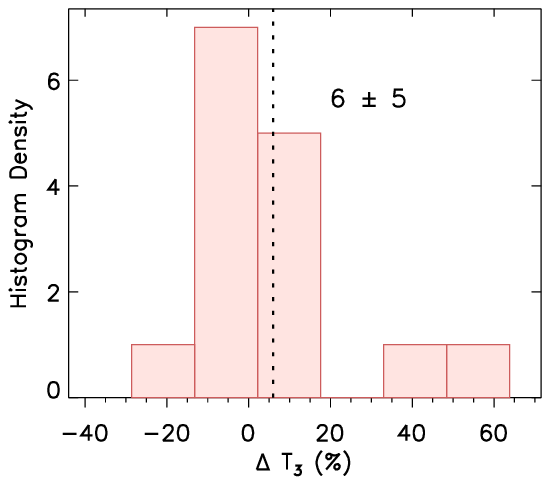}
\includegraphics[width=6.cm]{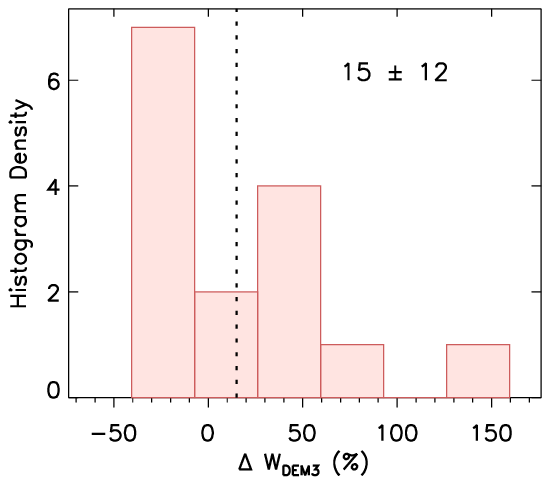}
\caption{Histograms of the percentage change in $EM$, $T,$ and $\wdem$ with respect to the value at P1.  The mean values and standard error are printed, and the dashed line shows the mean value.}
 \label{fig:windows_scatter}
\end{figure*}

\begin{figure}
\centering
\includegraphics[width=9.cm]{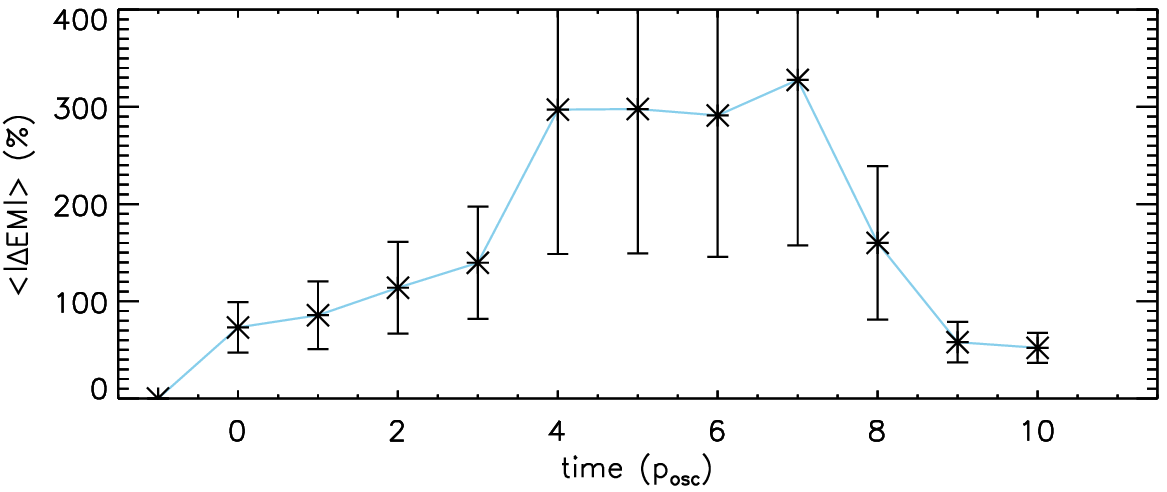}
\includegraphics[width=9.cm]{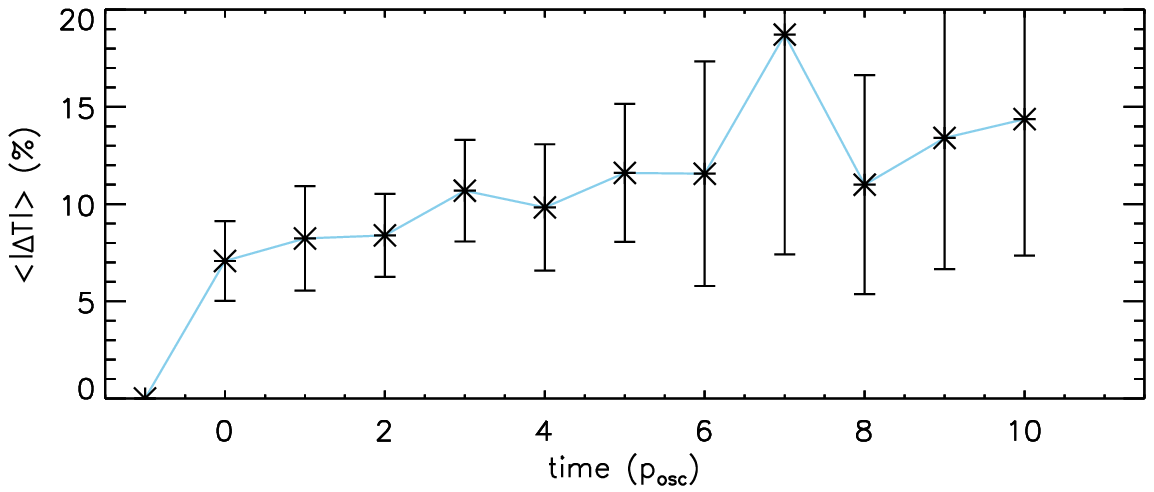}
\includegraphics[width=9.cm]{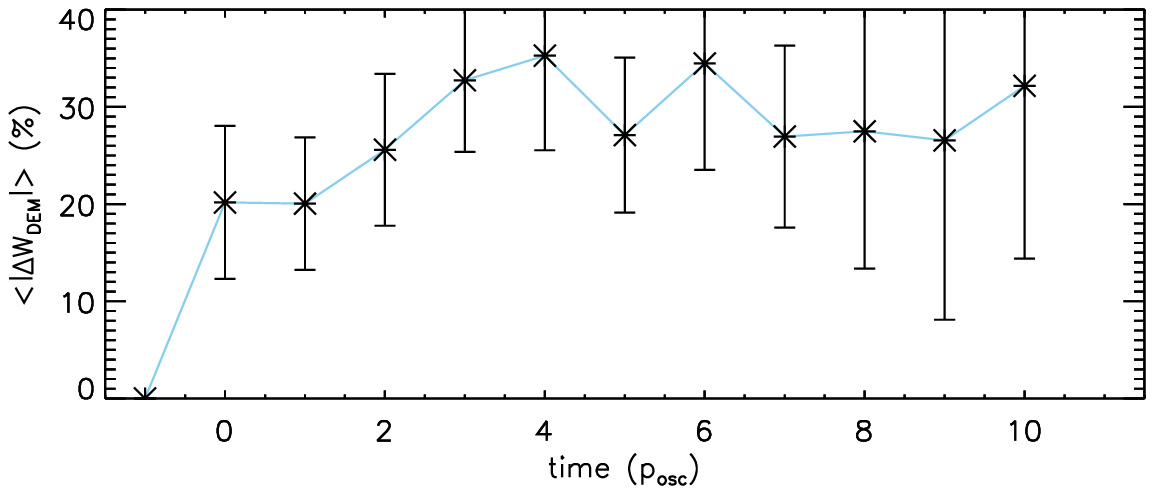}
\caption{Time series of the unsigned percentage variations of $EM$, $T,$ and $\wdem$ averaged over the oscillation period, and then over all events. The percentage change is with respect to the starting value, at -1 $\posc$ (before the initial perturbation). Zero represents the first oscillation period immediately following the initial perturbation of the loop ($t_0$). The time series extends until 10$\posc$ after $t_0$. }
 \label{fig:period_ave}
\end{figure}

\begin{table}[]
\caption{Average parameters during the time intervals P1, P2, and P3 over all events. The top three rows show the average value of the parameters, as seen in Fig.~\ref{fig:windows_para}. The middle three rows give the average percentage change with respect to the values in P1. The bottom three rows are the same as the middle three, but the average unsigned percentage changes are given. All uncertainties are the standard error on the associated average.}
\begin{tabular}{c|c|c|c}
\hline                                         
& P1           & P2           & P3          \\ \hline
$\langle EM\rangle \,(10^{26} \,\mathrm{cm}^{-5})$          & 0.7 $\pm$ 0.3   & 1.0 $\pm$ 0.4   & 1.0 $\pm$ 0.4     \\
$\langle T\rangle \,$(MK)                & 0.98 $\pm$ 0.04 & 1.00 $\pm$ 0.06 & 1.04 $\pm$ 0.07    \\
$\langle W_{DEM}\rangle \,$(MK)          & 0.60 $\pm$ 0.05 & 0.61 $\pm$ 0.07 & 0.65 $\pm$ 0.07   \\
$\langle \Delta EM\rangle \,$(\%)        & -               & 110 $\pm$ 60     & 130 $\pm$ 60       \\
$\langle \Delta T\rangle \,$(\%)         & -               & 2 $\pm$ 4      & 6 $\pm$ 5        \\
$\langle \Delta W_{DEM}\rangle \,$(\%)   & -               & 7 $\pm$ 10       & 15 $\pm$ 12        \\
$\langle |\Delta EM|\rangle \,$(\%)      & -               & 140 $\pm$ 50      & 160 $\pm$ 50     \\
$\langle |\Delta T|\rangle \,$(\%)       & -               & 9 $\pm$ 3      & 12 $\pm$ 4        \\
$\langle |\Delta W_{DEM}|\rangle \,$(\%) & -               & 26 $\pm$ 8      & 33 $\pm$ 9       
\\
\hline
\end{tabular}
\label{tab:2}
\end{table}

We now present a collective analysis of the behaviour of $EM$, $T,$ and $\wdem$ averaged during phases P1, P2, and P3, defined using the corresponding damping time. Additionally, we consider these parameters averaged over the corresponding oscillation period ($\posc$) from one period before the initial perturbation to ${10\, \posc}$ afterwards. The uncertainties presented are the standard error on the mean from averaging over all events.

\subsection{Averaging in three time intervals}

For each event the time intervals P1, P2, and P3 were calculated as described in Section~\ref{sect:init_process}. All 171 $\AA$ TD maps are shown in Appendix~\ref{app:1} with the intervals overplotted. The parameters $EM$, $T,$ and $\wdem$ were then averaged during these intervals for each event. Figure~\ref{fig:windows_para} shows histograms of these values, with the mean values and associated standard error indicated. These mean values over the 15 events are also given in the top three rows of Table~\ref{tab:2}. The mean values all remain in error between the three time windows. No correlations were found between the values of the $EM$, $T,$ and $\wdem$ and the period, loop length, or damping time. The mean temperature is 1 MK, which is expected because we analysed warms loops that are clearly visible at 171 $\AA$. The mean values of $EM$ and $\wdem$ are $0.7\times 10^{26}\mathrm{cm}^{-5}$ and $0.6$ MK, respectively.

Next we studied the percentage variations of the parameters with respect to the value in P1, which allows for a collective analysis of all the events. For example, ${\Delta T_2 = 100\times (T_2 -T_1)/T_1}$ gives the percentage change of the temperature from P1 to P2. The mean was then taken over all events, giving ${\langle \Delta T_2 \rangle}$, or ${\langle |\Delta T_2| \rangle}$ when we take the mean of the unsigned variations. 

The individual values for each event are plotted as histograms in Fig.~\ref{fig:windows_scatter}, with the mean values indicated. There is a clear tendency towards large increases in the $EM$ for some events, reflected in the mean percentage variations of 110\%\ and 130 $\%$. This is likely due to cases where some part of the eruptive plasma (or other perturbed loop) passes through the observational slit. There are no clear trends towards positive or negative variations for $T$ or $\wdem$. The mean variations for P2 and P3 with respect to P1 are also given in rows 3--6 of Table~\ref{tab:2}.

The mean unsigned variations are given in the bottom three rows of the table; we do not present the associated histograms. There is a $\approx$ 150\%, 15\%,\ and $\approx 30\%$ variation from P1 to P2 in the $EM$, $T,$ and $\wdem$, respectively, with little further variation between P2 and P3. This indicates that the eruption and initial perturbation of the loop account for most of this variation, or the first few cycles of oscillation. 

\subsection{Averaging over oscillation cycles}
To specifically investigate changes over shorter timescales we then averaged the time series over the oscillation period, $\posc$. This was done from one period before $t_0$ ($-1 \posc$) until ten periods after ($10\, \posc$) for each event, defining the percentage variation with respect to the value at $-1 \posc$. The mean value over all events was then computed for each time step, again defining the standard error, so that we have ${\langle \Delta EM \rangle}$, ${\langle \Delta T \rangle}$, and ${\langle \Delta \wdem \rangle}$ as a function of time. For many events, $10\, \posc$ extends beyond the data range, so that later time steps are averages over fewer events. This caveat should be recalled when the trends beyond 3-4 $\posc$ are discussed. We also note that in most cases, the oscillation has largely damped after a few periods, but we continued the period-averaged time series where possible regardless of this.

The time series averaged over all events are shown in Fig.~\ref{fig:period_ave}. Zero represents 1~$\posc$ immediately following the initial perturbation of the loop ($t_0$). The signed variations are not presented because they are largely within the error of zero. All three parameters undergo some variations from $-1 \posc$ to $0$, $\approx 80\%, 7\%,$ and 20 $\%$ for ${\langle |\Delta EM| \rangle}$, ${\langle |\Delta T| \rangle,}$ and ${\langle |\Delta \wdem| \rangle}$, respectively. This corresponds to changes during the initial displacement of the loop. Over the next few oscillation periods, there is little further change in the average variation with respect to the starting values. ${\langle |\Delta EM| \rangle}$ shows a peak between 4-8 $\posc$, supporting the other indications that the $EM$ generally shows a strong variation some time after the eruption and initial perturbation of the loop, which is not seen in the other parameters.

In summary, we find that $EM$, $T,$ and $\wdem$ all vary at the time of the initial loop displacement, but there is no tendency towards an increase or decrease across all events. Therefore the variation from P1 to P2 in the previous section is likely due to the displacement of the loop, rather than the oscillation. The $EM$ time series also shows large variations after a few oscillation periods.
\section{Discussion}
\label{sect:disc}

Our sample of oscillating coronal loops has an average loop length of 300 Mm, a kink mode period of 9 minutes, and an exponential damping time of 18 minutes. These are in line with the typical values of the distributions in \cite{2019ApJS..241...31N}. The oscillating loops (or individual loop threads) were selected based on observations at 171 $\AA$ and are typically not prominent at other wavelengths. They are therefore typically only well seen in the DEM maps at temperatures of about 0.7--1.2 MK, with some exceptions. The mean temperature value is 1 MK. The mean values of $EM$ and $\wdem$ are $0.7\times 10^{26}\mathrm{cm}^{-5}$ and $0.6$ MK, respectively. It is important to reiterate that we only calculated these parameters for the central 5 pixels of the TD maps. Therefore the measurements are largely comprised of emission from the loop  core, the position of which was determined at 171 $\AA$.

Two cases studies were presented in Section~\ref{sect:case}. For both we observed changes in the loop parameters before the time interval of interest as well as during it. The first example was dominated by additional plasma from the eruption seen at 94 and 131 $\AA$, whereas the second did not show any sharp changes due to the eruption. These two situations were present in the 15 events we analysed. We see evidence for periodic modulation of the parameters during the oscillation in some cases, and further in-depth study may help characterise to what extent this is due to the variation of the column depth, background structures, or other effects \citep[e.g.][]{2003A&A...397..765C,2016ApJS..223...23Y, 2016ApJS..223...24Y}. Loop width and intensity variations associated with the oscillation have been noted in other observational studies \citep[e.g.][]{2011ApJ...736..102A,2012ApJ...751L..27W}, and a relation to density variations is postulated. In the former observation, it was noted that the measured intensity modulation would require an LOS angle change, which is incompatible with the observed oscillation, but this does not rule out the influence of other structures along the LOS. We did not verify this here, and it requires a careful event-by-event analysis.

There are no significant shifts in the mean values of the emission measure, temperature, or DEM width over P1--P3; additionally, there are no clear trends in the signed percentage changes to positive or negative variations, except for the emission measure. On average, we detect $\approx 150$\%, 15\%,\ and 30$\%$ unsigned variations from P1 to P2 in $EM$, $T,$ and $\wdem$, respectively. This does not significantly change for P3. Based on this, we conclude that the main detected changes in emission measure (and density by proxy) and temperature occur following the initial perturbation. Shorter timescale variations were analysed by averaging the time series over the oscillation period for each event. It is found that the parameters undergo large average changes during the eruption, emphasising the above results and further localising the main variations to the initial perturbation. However, the mean unsigned variation of the emission measure shows a peak after a few oscillation cycles as a result of the cases where additional plasma from the eruption, or neighbouring loops, passes the observational slit. This is linked to the intensity peaks at 94 and 131 \AA~shown in Fig.~\ref{fig:case1}, which appear to be related to multi-thermal plasma from the eruption and flare in the vicinity of the TD map slit. This is apparent for several of the studied events and may therefore support this conclusion. This may still be related to a change in the background conditions, and may be important for the analysis of the loops and their oscillations.

 \cite{2019ASSL..458.....A} pointed out that coronal loops at temperatures of 1--2 MK are expected to undergo efficient radiative cooling. Consequently, the observed lifetime of a loop seen at 171 or 193 \AA~should be 10–20 min in the absence of heating, but the apparently monolithic loop threads observed here generally last one to two hours, although they do visibly evolve in this time (see Appendix~\ref{app:1}). Further analysis is required to determine to what extent the detected evolution is due to general thermal cycles of the loop bundles (as mentioned in the introduction) or to thermodynamic changes in the active region due to the eruption.

\subsection{Evidence for KHI?}
One motivation for this work was to make a first step in searching for evidence of KHI occurring for oscillating coronal loops. Signatures such as a varying intensity ratio between hotter and cooler channels, the appearance of substructure, and a broadening of the loops DEM have been predicted \citep{2017ApJ...836..219A, 2018ApJ...863..167G, 2018A&A...620A..65V}.

The mean value of $\wdem$ is measured to increase in P2 and P3, but the values are all within the error. In Table~\ref{tab:2} and Fig.~\ref{fig:period_ave} we show that the emission measure, temperature, and DEM width do vary during the oscillation, but the case studies showed that this is likely part of a longer timescale evolution. Sharper changes also occur at and during the initial displacement of the loop.

\cite{2018ApJ...853...35T} showed that for loops with large inhomogeneous layers, which seem to be more common \citep{2017A&A...605A..65G}, the effect of twist may be important in suppressing the development of shear instabilities at the loop boundary. Loops being in an initial state with a large inhomogeneous layer also delay the onset of KHI, and these effects may help explain the lack of hard observational evidence found here. However, we emphasise that we only analysed the centre of loops as seen at 171 \AA, and might therefore miss predicted changes at the loop edges or trends in loops that are better seen at other AIA wavelengths. However, the TD maps in Figs.~\ref{fig:alltd1} and \ref{fig:alltd2} show the generally large variability of the loop structure and intensity over the timescales of the oscillation. This should be the subject of a more detailed study to determine if it results from non-linear processes.

\subsection{Implications for seismology}

Multiple theoretical studies have analysed the effect of the varying loop structure on the oscillations they exhibit. If the loops vary strongly during the oscillation, for example in density, then it is clear that this will affect the analysis of the oscillation and seismology. This analysis assumes that the loops are in a steady state. Making a first attempt at quantifying the evolution of oscillating loops observationally was another aim of this study. 

Details of the detected variation have been discussed extensively above. To study the oscillations, the variation of the loop after the initial perturbation is most important. During a time period that exceeds the oscillation period following the eruption ten times, we find that the $EM$, $T,$ and $\wdem$ typically vary by about 100\%, 10\%,\ and 35 \% respectively, compared to the values prior to the perturbation. However, this may still be strongly influenced by the limited background subtraction.

Variation of the loop interior density would vary the associated Alfv\'en speed and therefore the kink speed and observed period of oscillation. This potential source of uncertainty would already be taken into account by the error on the estimated period, which should reflect its variation, and therefore on any seismologically determined parameters. However, this ignores the interplay between density, temperature, and magnetic field in maintaining pressure balance across the loop. Additionally, the transverse structure of the loop may evolve, which determines the damping behaviour through resonant absorption. Here, we note that the measured loop properties are seen to evolve on timescales that are important for the oscillations, and should be the subject of further investigation.

\subsection{Limitations}
Our preliminary analysis has various limitations that may need to be addressed in further work. For example, we did not analyse the evolution of the background plasma, but the density contrast is also important in the context of waves and seismology. We did not try to model the transverse density profiles of the loops either, as performed in \cite{2017A&A...600L...7P} and \cite{2017A&A...605A..65G, 2018ApJ...863..167G}. This would require careful treatment of the background in all wavelengths, and introduce the further complexity of applying it to all structures over time.  This may be a natural extension of the work in the future, however. 

Our results can clearly be heavily influenced by the multiple overlapping structures with different temperatures (see Fig.~\ref{fig:DEM_EM_temp} for an example). These structures often do not appear to be part of the same monolithic loop, but separate structures at different temperatures. Active regions appear more populated at higher temperatures, and this may mask the actual higher temperature component of the loop of interest \citep{2014ApJ...795..171S, 2016ApJ...831..199S}.

Recently, efforts have increased to improve DEM analysis codes \citep[e.g.][]{2019SoPh..294..135M} and forward-modelling approaches \citep[e.g.][]{2019ApJ...884...43P}. These techniques should be considered for future studies, with multi-wavelength analysis at multiple positions along the oscillating loop.

\subsection{Conclusions}

Following a preliminary analysis, we find that the emission measure, temperature, and DEM distribution width of coronal loops undergo significant variation on timescales relevant for the study of their oscillation. This is expected because of the variability seen in the 171 $\AA$ TD maps. There is no clear trend towards increases or decreases, however, and so the mean values of the distributions do not change significantly. The emission measure seems to be the most sensitive parameter to the plasma from the eruption and often shows significant variation. Furthermore, we have shown that most of the variation that is not due to eruptive plasma passing through the slit occurs at the time of the initial perturbation of the loop, likely related to the change in the column depth, the background structures, or genuine perturbation of the thermodynamic equilibrium. Through this publication we indicate a sample of 15 high-quality oscillation events and their characterising parameters for further analysis by the community. 

\begin{acknowledgements}
CRG is supported by ERC Synergy grant WHOLE SUN 810218. GN acknowledges the support of the CGAUSS project by the German Aerospace Centre (DLR) under grant 50OL1901 at the University of Göttingen and the "Rita-Levi-Montalcini 2017" fellowship funded by the Italian Ministry of Education, University and Research at the University of Calabria. The data is used courtesy of the AIA team. The DEM inversion code was used courtesy of I. G. Hannah and E. P. Kontar.
\end{acknowledgements}

\bibliographystyle{aa} 
\bibliography{ref.bib} 

\begin{appendix} 
\section{Time-distance maps}
\label{app:1}

\begin{figure}
\centering
\includegraphics[width=9.cm]{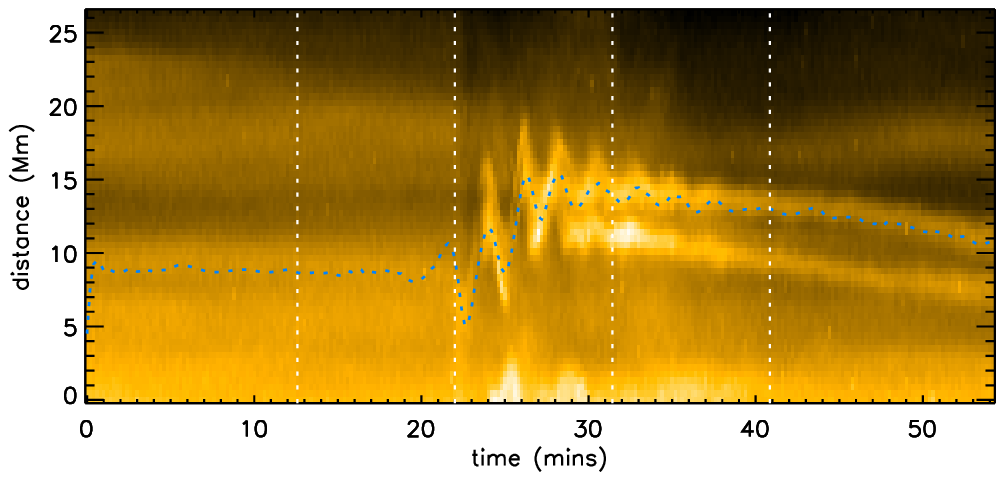}
\includegraphics[width=9.cm]{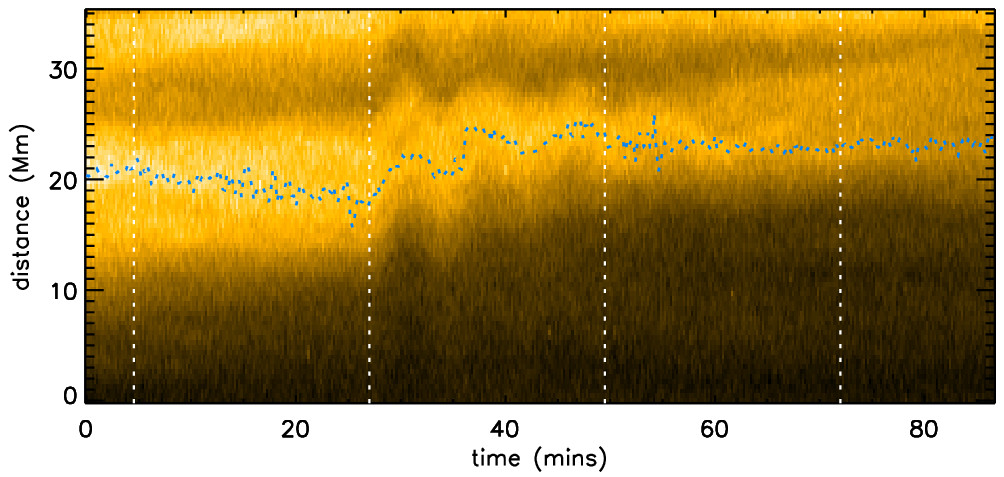}
\includegraphics[width=9.cm]{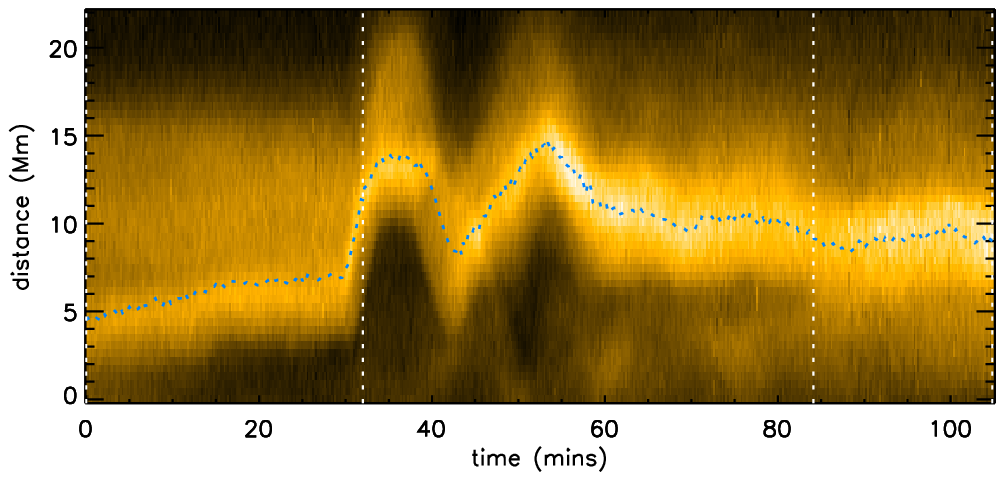}
\includegraphics[width=9.cm]{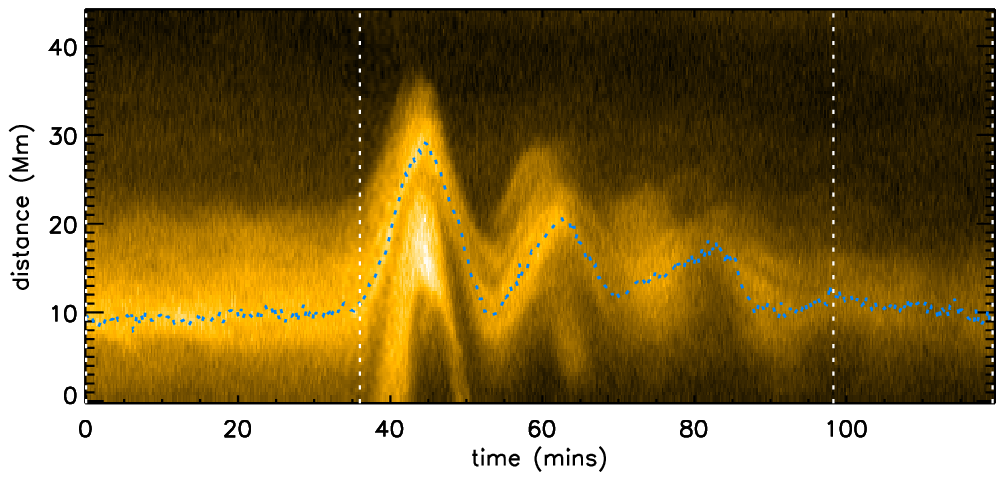}
\includegraphics[width=9.cm]{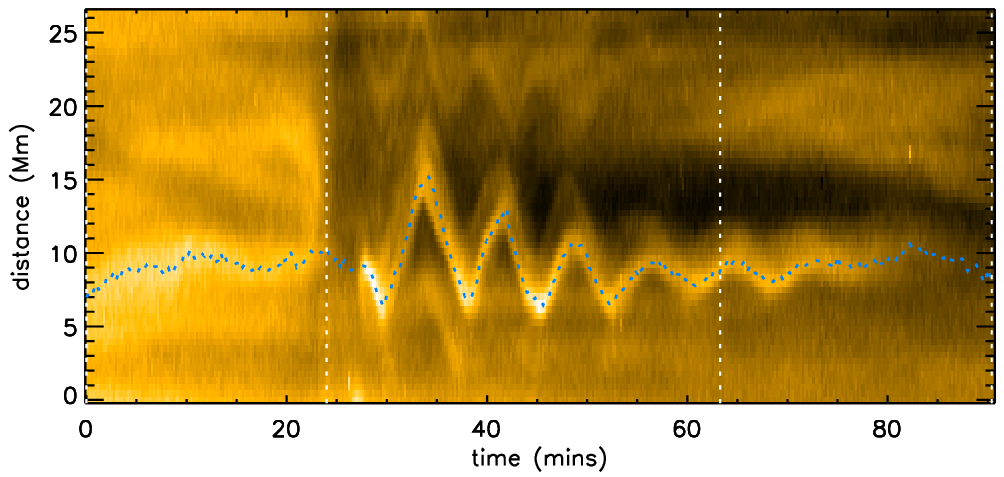}
\caption{TD maps of the analysed oscillations (except for the two case studies). In the same order as Table~\ref{tab:1}, top to bottom.}
 \label{fig:alltd1}
\end{figure}

\begin{figure*}
\centering
\includegraphics[width=9.cm]{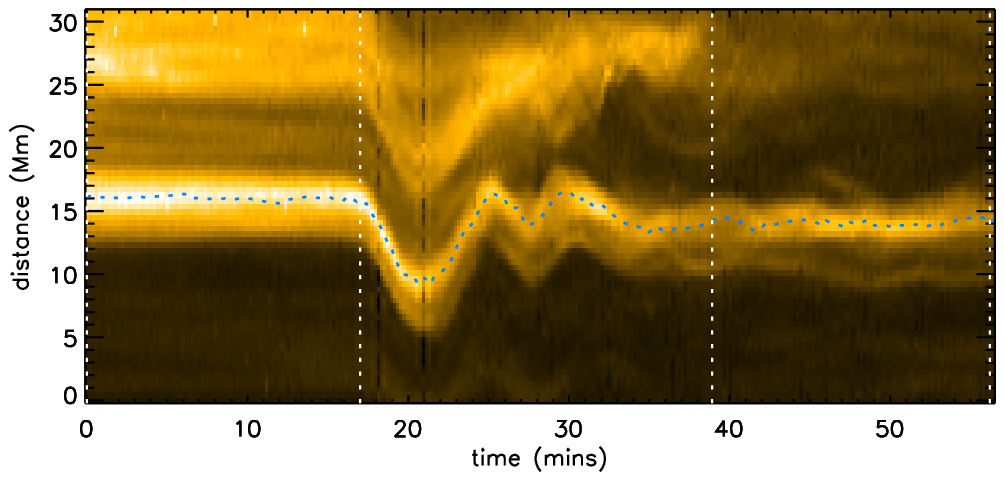}
\includegraphics[width=9.cm]{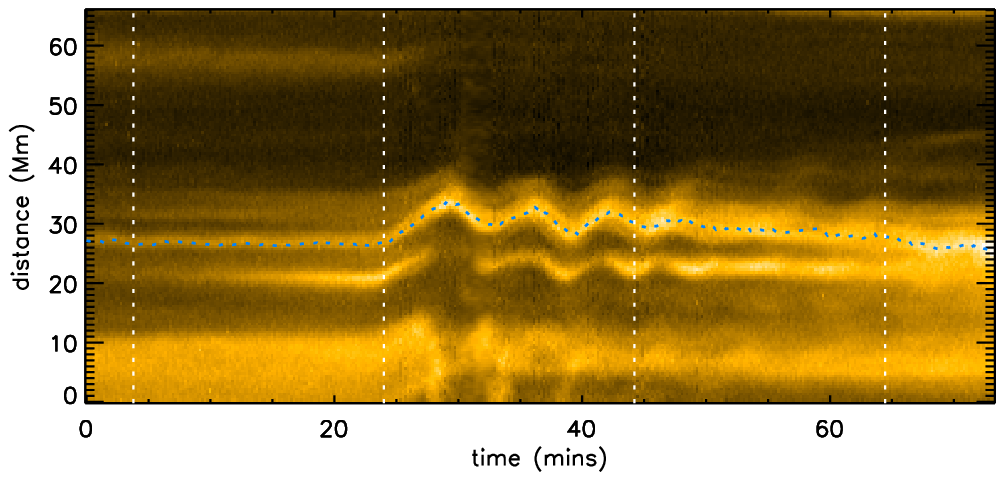}
\includegraphics[width=9.cm]{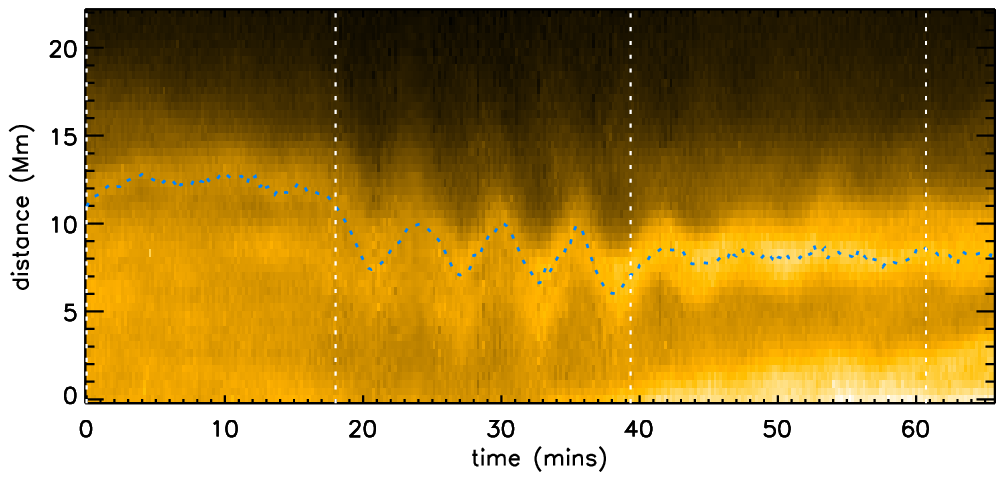}
\includegraphics[width=9.cm]{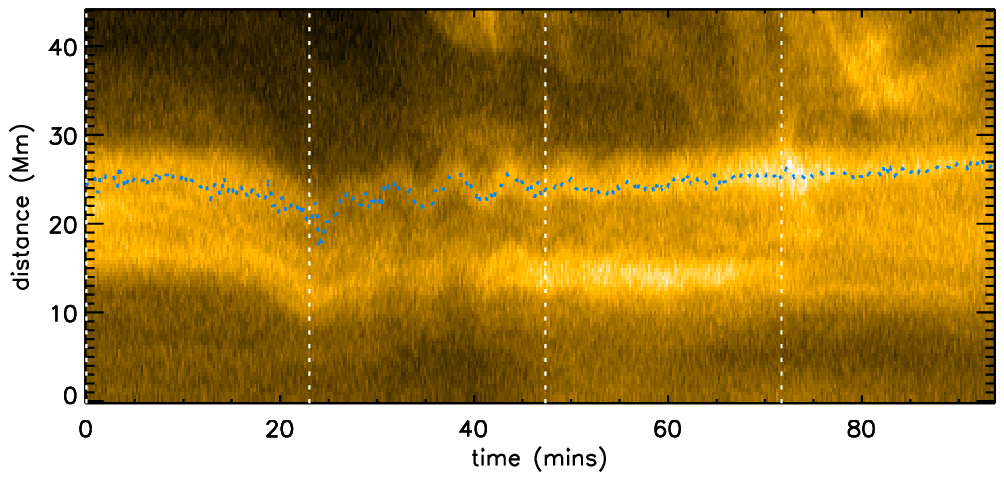}
\includegraphics[width=9.cm]{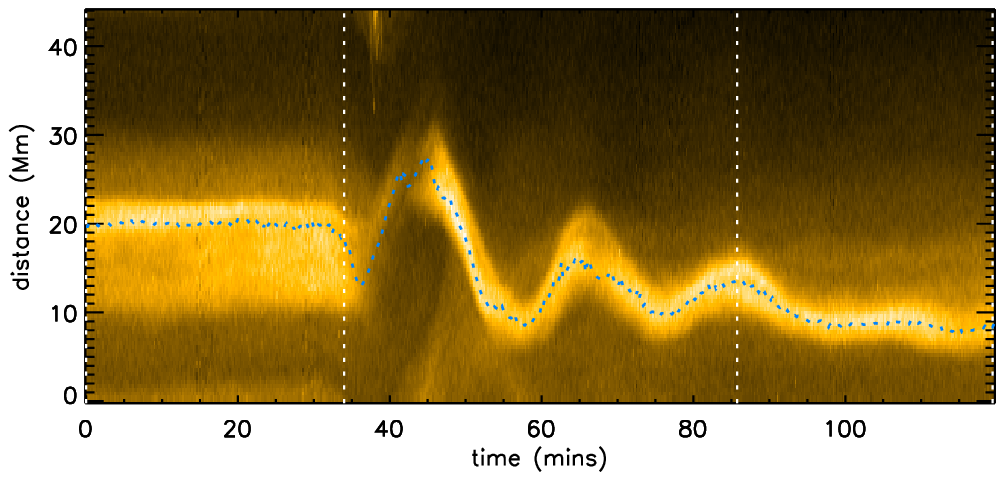}
\includegraphics[width=9.cm]{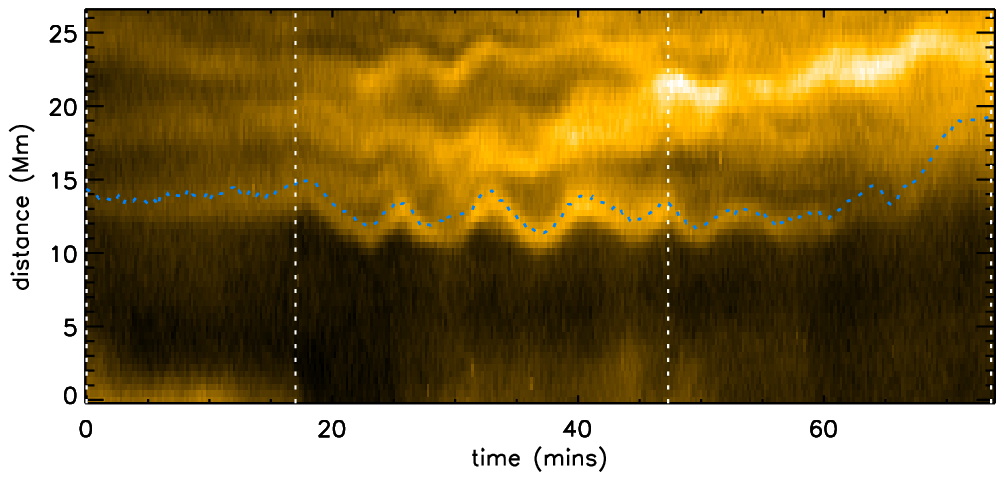}
\includegraphics[width=9.cm]{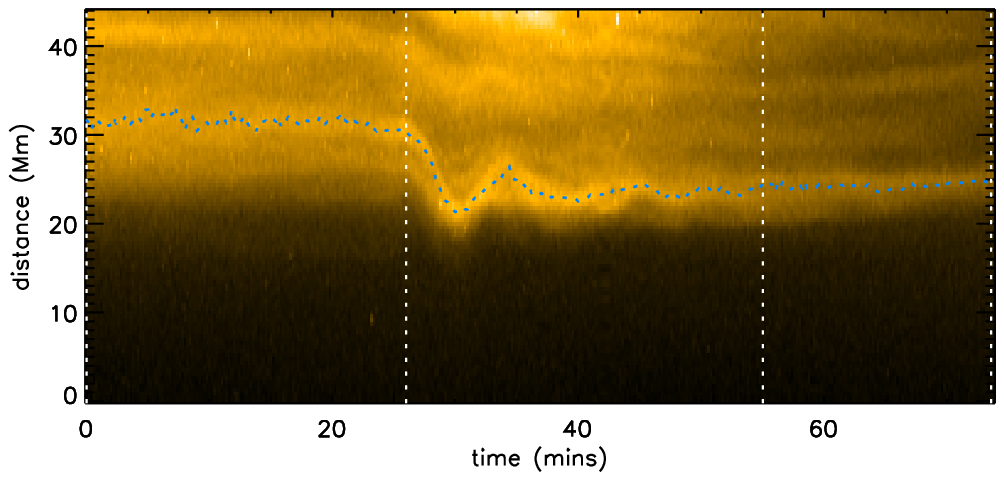}
\includegraphics[width=9.cm]{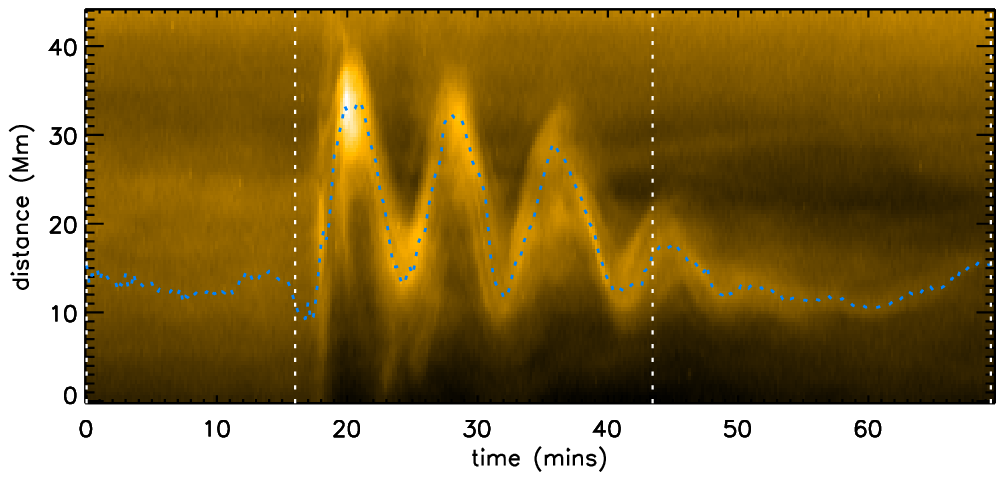}
\caption{TD maps of the analysed oscillations (except for the two case studies). In the same order as Table~\ref{tab:1}, top to bottom, left column first.}
 \label{fig:alltd2}
\end{figure*}

We present all of the TD maps at 171 \AA~for the analysed loops in Figs.~\ref{fig:alltd1} and~\ref{fig:alltd2} (except for the two case studies). Overplotted are the determined loop centre positions that we used to  detrend~the TD maps at all wavelengths. The parameters presented throughout the paper are averaged over +/- 2 pixels about this centre position. Vertical dashed lines depict the time intervals P1, P2, and P3, as described in Section~\ref{sect:init_process}.

All of the oscillations analysed are associated with an eruption, which directly perturbs the loop in most cases \citep{2015A&A...577A...4Z}. Several of the oscillations appear to remain at a constant amplitude for several oscillation cycles, or even increase in amplitude, before eventually damping (e.g. the top right three panels of Fig.~\ref{fig:alltd2}). Some oscillations also have a non-sinusoidal shape, for example the third panel from top in Fig.~\ref{fig:alltd1}. Finally, the quality factor of the oscillations varies widely, from two or three oscillation cycles up to eight or nine (which may then be followed by a low-amplitude decayless component).

Finally, most of the loops appear to vary in structure and intensity over the course of the observations. Taken together, all of these different features present opportunities for further detailed study, which is beyond the scope of the present work. These maps also show the difficulty of employing a more rigorous background subtraction because contaminating structures are visible in the vicinity of the oscillating loops at 171 \AA. Typically, the loops are then not clearly seen at the hotter wavelengths, which prevents a consistent subtraction at all wavelengths in most cases.

\end{appendix}
\end{document}